\documentclass[journal,twoside,letterpaper]{IEEEtran}

\usepackage{amsmath,amsfonts}
\usepackage{algorithmic}
\usepackage{array}
\usepackage{textcomp}
\usepackage{stfloats}
\usepackage{url}
\usepackage{verbatim}
\usepackage{graphicx}
\def\BibTeX{{\rm B\kern-.05em{\sc i\kern-.025em b}\kern-.08em
    T\kern-.1667em\lower.7ex\hbox{E}\kern-.125emX}}
\usepackage{balance}

\usepackage{amssymb}
\usepackage{multirow}
\usepackage{svg}
\usepackage{booktabs}
\usepackage[font=footnotesize]{caption}
\usepackage{subcaption}
\usepackage{xcolor}

\usepackage[hidelinks]{hyperref}
\usepackage{orcidlink}

\usepackage[
    style=ieee,   
    ]{biblatex}
\usepackage{./resources/tikz/tikzit}

\tikzstyle{Rectangle Medium}=[fill={rgb,255: red,42; green,71; blue,101}, draw=black, shape=rectangle, minimum width=2cm, text=white, minimum height=0.75cm, align=center]
\tikzstyle{Rectangle Large}=[fill={rgb,255: red,42; green,71; blue,101}, draw=black, shape=rectangle, text=white, minimum width=4cm, minimum height=2cm, align=center]
\tikzstyle{Red Text}=[fill=none, draw=none, shape=rectangle, tikzit draw={rgb,255: red,255; green,131; blue,131}, tikzit fill={rgb,255: red,255; green,131; blue,131}, text={{rgb,255: red,255; green,131; blue,131}}, align=center]
\tikzstyle{Small Circle}=[fill={rgb,255: red,42; green,71; blue,101}, draw=black, shape=circle, text=white]
\tikzstyle{Text}=[fill=none, draw=none, shape=rectangle, tikzit draw=black, align=center]
\tikzstyle{White Text}=[fill=none, draw=none, shape=rectangle, tikzit draw=black, align=center, text=white]
\tikzstyle{Rectangle Small}=[fill={rgb,255: red,42; green,71; blue,101}, draw=black, shape=rectangle, text=white, align=center, minimum height=0.5cm, minimum width=1.5cm]
\tikzstyle{Rectangle MedLarge}=[fill={rgb,255: red,42; green,71; blue,101}, draw=black, shape=rectangle, text=white, align=center, minimum height=1.33cm, minimum width=2.5cm]
\tikzstyle{Small Outline Circle}=[fill=white, draw=black, shape=circle, inner sep=0pt]

\tikzstyle{Arrow}=[draw=black, ->, line width=.025cm]
\tikzstyle{Edge}=[-, draw=black, line width=0.025cm, fill=none]
\tikzstyle{Red Dashed}=[-, draw={rgb,255: red,255; green,131; blue,131}, dashed, line width=0.025cm]
\tikzstyle{Dashed}=[-, draw={rgb,255: red,136; green,136; blue,136}, dashed, line width=0.025cm]
\tikzstyle{Freebox}=[-, draw=black, fill={rgb,255: red,42; green,71; blue,101}]
\tikzstyle{Freebox Green}=[-, draw=black, fill={rgb,255: red,115; green,167; blue,144}]
\tikzstyle{Freebox Yellow}=[-, draw=black, fill={rgb,255: red,215; green,177; blue,124}]
\tikzstyle{Freebox Red}=[-, draw=black, fill={rgb,255: red,255; green,131; blue,131}]
\tikzstyle{Thick Dashed}=[-, line width=0.025cm, draw=black, dashed]
\tikzstyle{Black Lines}=[-, fill=none, pattern=north east lines, pattern color={{rgb,255: red,42; green,71; blue,101}}, draw=black, tikzit fill={rgb,255: red,191; green,191; blue,191}, tikzit draw={rgb,255: red,241; green,239; blue,228}, line width=0.025cm]
\tikzstyle{Blue Background}=[-, fill={{rgb,255: red,42; green,71; blue,101}}, fill opacity=0.3, draw=none]
\tikzstyle{Green Background}=[-, fill={{rgb,255: red,115; green,167; blue,144}}, fill opacity=0.45, draw=none]
\tikzstyle{Black Lines Dashed}=[-, fill=none, pattern=north east lines, pattern color={{rgb,255: red,42; green,71; blue,101}}, draw=black, tikzit fill={rgb,255: red,191; green,191; blue,191}, tikzit draw={rgb,255: red,241; green,239; blue,228}, line width=0.025cm, dashed]
\tikzstyle{Blue Dots}=[-, fill=none, pattern=horizontal lines, pattern color={{rgb,255: red,215; green,177; blue,124}}, tikzit fill={rgb,255: red,42; green,71; blue,101}, line width=0.025cm]
\tikzstyle{Red Bricks}=[-, fill=none, pattern=crosshatch, line width=0.025cm, tikzit fill={rgb,255: red,255; green,131; blue,131}, pattern color={{rgb,255: red,115; green,167; blue,144}}]
\tikzstyle{Black Cross}=[-, fill=none, pattern=crosshatch dots, line width=0.025cm, tikzit fill={rgb,255: red,255; green,131; blue,131}, pattern color={{rgb,255: red,42; green,71; blue,101}}]
\tikzstyle{Red Bricks Dashed}=[-, fill=none, pattern=crosshatch, line width=0.025cm, tikzit fill={rgb,255: red,255; green,131; blue,131}, pattern color={{rgb,255: red,115; green,167; blue,144}}, dashed]
\tikzstyle{Dashed Arrow}=[->, line width=0.025cm, draw=black, dashed]
\tikzstyle{Thin Dashed Arrow}=[->, line width=0.025cm, dashed, draw={rgb,255: red,136; green,136; blue,136}]
\tikzstyle{Blue EC F}=[-, draw={rgb,255: red,42; green,71; blue,101}, fill={rgb,255: red,42; green,71; blue,101}, fill opacity=0.33, line width=0.25mm]
\tikzstyle{Green EC F}=[-, draw={rgb,255: red,115; green,167; blue,144}, fill={rgb,255: red,115; green,167; blue,144}, fill opacity=0.33, line width=0.25mm]
\tikzstyle{Yellow EC F}=[-, draw={rgb,255: red,215; green,177; blue,124}, fill={rgb,255: red,215; green,177; blue,124}, fill opacity=0.33, line width=0.25mm]
\tikzstyle{Pink EC F}=[-, draw={rgb,255: red,234; green,186; blue,185}, fill={rgb,255: red,234; green,186; blue,185}, fill opacity=0.33, line width=0.25mm]

\DeclareUnicodeCharacter{2212}{-}

\newcommand{\Description}[1]{}

\newcommand\copyrighttext{%
  This work has been submitted for possible publication. Copyright may be transferred without notice, after which this version may no longer be accessible.
}

\newcommand\copyrightnotice{%
\begin{tikzpicture}[remember picture,overlay]
\node[anchor=south,yshift=5pt] at (current page.south) {\fbox{\parbox{\dimexpr\textwidth-\fboxsep-\fboxrule\relax}{\copyrighttext}}};
\end{tikzpicture}%
}

\begin{document}
\title{Workload Buoyancy: Keeping Apps Afloat by Identifying Shared Resource Bottlenecks}

\author{Oliver~Larsson\orcidlink{0000-0003-0395-8313},
    Thijs~Metsch\orcidlink{0000-0003-3495-3646}, 
    Cristian~Klein\orcidlink{0000-0003-0106-3049}, 
    Erik~Elmroth\orcidlink{0000-0002-2633-6798}
\thanks{
    Funding for this project was provided
    by the eSSENCE Programme under the
    Swedish Government’s Strategic Research Initiative.
}%
\thanks{Oliver Larsson, Cristian Klein, and Erik Elmroth are with the 
Department of Computing Science, Ume\aa{} University, SE-90187 Ume\aa, Sweden 
(e-mail: olars@cs.umu.se; cklein@cs.umu.se; elmroth@cs.umu.se).}%
\thanks{Thijs Metsch is unaffiliated, Germany 
(e-mail: tmetsch@engjoy.eu).}%
}

\markboth{}{Larsson \MakeLowercase{\textit{et al.}}: Workload Buoyancy: Keeping Apps Afloat by Identifying Shared Resource Bottlenecks}%

\maketitle

\copyrightnotice

\begin{abstract}
  Modern multi-tenant, hardware-heterogeneous computing environments 
pose significant challenges for effective workload orchestration. 
Simple heuristics for assessing workload performance, such as CPU utilization or 
application-level metrics, are often insufficient to capture the complex 
performance dynamics arising from resource contention and 
noisy-neighbor effects. In such environments, 
performance bottlenecks may emerge in any shared system resource, 
leading to unexpected and difficult-to-diagnose degradation.

This paper introduces buoyancy, a novel abstraction for characterizing 
workload performance in multi-tenant systems. Unlike traditional approaches, 
buoyancy integrates application-level metrics with system-level insights of 
shared resource contention to provide a holistic view of performance dynamics. 
By explicitly capturing bottlenecks and headroom across multiple resources, 
buoyancy facilitates resource-aware and application-aware orchestration in a 
manner that is intuitive, extensible, and generalizable across heterogeneous 
platforms.
We evaluate buoyancy using representative multi-tenant workloads to 
illustrate its ability to expose performance-limiting resource interactions.
Buoyancy provides a 19.3\% better indication of bottlenecks compared
to traditional heuristics on average.
We additionally show how buoyancy can act as a drop-in replacement for 
conventional performance metrics, enabling improved observability 
and more informed scheduling and optimization decisions.

\end{abstract}


\section{Introduction}

\IEEEPARstart{I}{mproving} efficiency and hardware utilization has always been one of the primary
objectives of compute cluster management~\cite{verma_large-scale_2015, zhang_zeus_2021}.
Through the use of virtualization, resource overcommitment, and scheduling 
techniques, cloud providers can increase the overall utilization and cost 
efficiency of their clusters. 
While beneficial to their bottom line, the noisy neighbor effects caused by the
additional resource congestion created may prove problematic for application owners 
seeking to meet their service-level objectives (SLO)~\cite{lorido-botran_unsupervised_2017}, who in turn
are required to increase their resource requests to counteract.

Such SLOs are typically expressed in terms 
of thresholds in key performance indicators (KPI) specific
to the application~\cite{silvaPerformanceEvaluationCloud2025}. 
Due to varying sensitivities to shared system resources, 
this application-level performance indicator is rarely immediately correlated with the 
allocation of traditionally assignable hardware resources such as CPU cores and memory~\cite{larssonHardwareLevelQoSEnforcement2025}.
Rather, a much more complex dynamic is at play, where congestion in other shared resources such as 
last-level cache (LLC) or memory bandwidth can be a major contributing factor to application-level performance issues~\cite{chenOLPartOnlineLearning2023}.
This problem is further complicated in the cloud by resource inconsistencies such as the 
performance of single CPU cores varying between 
generations and models~\cite{dravaiPerformanceEfficiencyMultigenerational2025}, cache configurations, NUMA topologies, 
and that such hardware traits are hidden in the virtualized hardware-heterogeneous environments.

This apparent complexity combined with recent trends towards intent-driven orchestration, where application owners
declare high-level objectives rather than low-level resource requests~\cite{nastic_sloc_2020,metsch_intent-driven_2023}, suggest
a need for a streamlined approach to resource management and observability that captures the complex dynamics
between shared resources and application-level performance without increasing the cognitive load on application 
owners or cluster administrators.

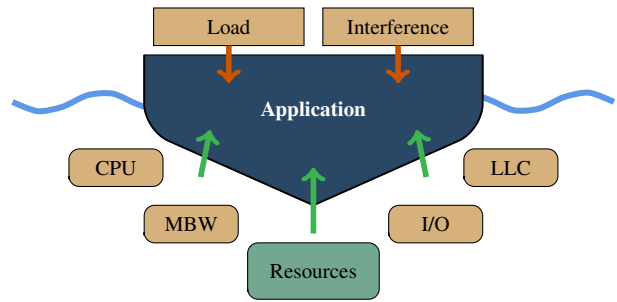
\begin{figure}[tb]
    \centering
    \footnotesize
    \begin{tikzpicture}
	\begin{pgfonlayer}{nodelayer}
		\node [style=none] (3) at (-4.5, -0.75) {};
		\node [style=none] (4) at (0, -2.75) {};
		\node [style=none] (5) at (4.5, -0.75) {};
		\node [style=none] (6) at (4.5, 1.25) {};
		\node [style=none] (7) at (-4.5, 1.25) {};
		\node [style=none] (9) at (-4.25, 2.5) {};
		\node [style=none] (10) at (-4.25, 1.5) {};
		\node [style=none] (11) at (-0.25, 1.5) {};
		\node [style=none] (12) at (-0.25, 2.5) {};
		\node [style=none] (13) at (0.25, 2.5) {};
		\node [style=none] (14) at (0.25, 1.5) {};
		\node [style=none] (15) at (4.25, 1.5) {};
		\node [style=none] (16) at (4.25, 2.5) {};
		\node [style=none] (19) at (-2.25, 1.5) {};
		\node [style=none] (20) at (-2.25, 0.5) {};
		\node [style=none] (21) at (2.25, 1.5) {};
		\node [style=none] (22) at (2.25, 0.5) {};
		\node [style=none] (23) at (-1.75, -3.75) {};
		\node [style=none] (24) at (-1.75, -5.25) {};
		\node [style=none] (25) at (1.75, -5.25) {};
		\node [style=none] (26) at (1.75, -3.75) {};
		\node [style=none] (28) at (-6.5, -1.25) {};
		\node [style=none] (29) at (-6.5, -2.25) {};
		\node [style=none] (30) at (-4, -2.25) {};
		\node [style=none] (31) at (-4, -1.25) {};
		\node [style=none] (33) at (-4.5, -2.75) {};
		\node [style=none] (34) at (-4.5, -3.75) {};
		\node [style=none] (35) at (-2, -3.75) {};
		\node [style=none] (36) at (-2, -2.75) {};
		\node [style=none] (38) at (4, -1.25) {};
		\node [style=none] (39) at (4, -2.25) {};
		\node [style=none] (40) at (6.5, -2.25) {};
		\node [style=none] (41) at (6.5, -1.25) {};
		\node [style=none] (43) at (-3, -2) {};
		\node [style=none] (44) at (-2.75, -0.75) {};
		\node [style=none] (45) at (0, -3.5) {};
		\node [style=none] (46) at (0, -1.75) {};
		\node [style=none] (47) at (3, -2) {};
		\node [style=none] (48) at (2.75, -0.75) {};
		\node [style=none] (49) at (2, -2.75) {};
		\node [style=none] (50) at (2, -3.75) {};
		\node [style=none] (51) at (4.5, -3.75) {};
		\node [style=none] (52) at (4.5, -2.75) {};
		\node [style=none] (54) at (-4.5, 0) {};
		\node [style=none] (56) at (-2.25, -1.75) {};
		\node [style=none] (57) at (2.25, -1.75) {};
		\node [style=none] (58) at (4.5, 0) {};
		\node [style=Text] (17) at (-2.25, 2) {Load};
		\node [style=Text] (18) at (2.25, 2) {Interference};
		\node [style=White Text] (8) at (0, -0.25) {\textbf{Application}};
		\node [style=Text] (32) at (-5.25, -1.75) {CPU};
		\node [style=Text] (42) at (5.25, -1.75) {LLC};
		\node [style=Text] (37) at (-3.25, -3.25) {MBW};
		\node [style=Text] (53) at (3.25, -3.25) {I/O};
		\node [style=Text] (27) at (0, -4.5) {Resources};
	\end{pgfonlayer}
	\begin{pgfonlayer}{edgelayer}
        \draw[draw={rgb,255: red,93; green,151; blue,231}, line width=.08cm, xshift=-8cm, scale=.509, domain=0:31.416, smooth, variable=\x] plot ({\x}, {sin(deg(\x) + 180) * .5});
		\draw [style=Freebox] (58.center)
			 to (6.center)
			 to (7.center)
			 to (54.center)
			 to (56.center)
			 to (4.center)
			 to (57.center)
			 to cycle;
		\draw [style=Freebox Yellow] (11.center)
			 to (12.center)
			 to (9.center)
			 to (10.center)
			 to cycle;
		\draw [style=Freebox Yellow] (15.center)
			 to (16.center)
			 to (13.center)
			 to (14.center)
			 to cycle;
		\draw [style=Arrow, line width=.075cm, draw={rgb,255: red,204; green,85; blue,0}] (19.center) to (20.center);
		\draw [style=Arrow, line width=.075cm, draw={rgb,255: red,204; green,85; blue,0}] (21.center) to (22.center);
		\draw [style=Freebox Green, rounded corners=4] (24.center)
			 to (25.center)
			 to (26.center)
			 to (23.center)
			 to cycle;
		\draw [style=Freebox Yellow, rounded corners=4] (29.center)
			 to (30.center)
			 to (31.center)
			 to (28.center)
			 to cycle;
		\draw [style=Freebox Yellow, rounded corners=4] (34.center)
			 to (35.center)
			 to (36.center)
			 to (33.center)
			 to cycle;
		\draw [style=Freebox Yellow, rounded corners=4] (39.center)
			 to (40.center)
			 to (41.center)
			 to (38.center)
			 to cycle;
		\draw [style=Freebox Yellow, rounded corners=4] (50.center)
			 to (51.center)
			 to (52.center)
			 to (49.center)
			 to cycle;
		\draw [style=Freebox, rounded corners=10, draw=none] (3.center)
			 to (4.center)
			 to (7.center)
			 to cycle;
		\draw [style=Edge, rounded corners=10] (7.center)
			 to (3.center)
			 to (4.center);
		\draw [style=Freebox, rounded corners=10, draw=none] (4.center)
			 to (5.center)
			 to (6.center)
			 to cycle;
		\draw [style=Edge, rounded corners=10] (4.center)
			 to (5.center)
			 to (6.center);
		\draw [style=Arrow, line width=.075cm, draw={rgb,255: red,60; green,181; blue,80}] (43.center) to (44.center);
		\draw [style=Arrow, line width=.075cm, draw={rgb,255: red,60; green,181; blue,80}] (45.center) to (46.center);
		\draw [style=Arrow, line width=.075cm, draw={rgb,255: red,60; green,181; blue,80}] (47.center) to (48.center);
	\end{pgfonlayer}
\end{tikzpicture}
    \caption{Illustration of the intuition behind the buoyancy concept. Here,
    the application is represented by a ship floating on a body of water. The
    goal is to keep the ship afloat, which is analogous to keeping the workload
    within its performance limits. Additional load or interference may cause the
    ship to sink, while adding resources increases the buoyancy of the ship.
    A ship with a greater buoyancy has a more margin and can withstand 
    larger increases in load or interference without sinking compared to a ship
    barely keeping above the waterline.} 
    \label{fig:buoyancy-illustration}
    \Description{A ship floating on water. The ship is labeled "Application" and the water is labeled ``Resources.'' 
    An arrow pointing down towards the ship is labeled ``Load/Interference'' 
    and an arrow pointing up from the ship is labeled ``Resources.''}
\end{figure}

To address this need, we introduce the concept of \emph{buoyancy}. An intuitive,
extensible, and generalizable way of characterizing workload performance
in dynamic, shared, and heterogeneous environments with complex system resource interdependencies. 
Buoyancy provides explicit indicators for bottlenecks and available headroom in individual resources, and combines these
with application-level performance metrics to provide a holistic view of both
workload and cluster performance. An illustration of the buoyancy concept is provided in Figure~\ref{fig:buoyancy-illustration}.

Specifically, we make the following contributions:
\begin{itemize} 
    \item We motivate the need for a practical and generalizable approach to 
        workload performance monitoring and resource management in shared, heterogeneous computing environments. 
        Our argument is grounded in the perspectives of key stakeholders, including application owners and cluster administrators.
    \item We introduce the concept of \emph{buoyancy}, an extensible and generalizable way of 
        characterizing workload performance that provides
        resource and application insights without prior profiling.
    \item We demonstrate the feasibility and benefits of workload buoyancy in
        a cloud native Kubernetes environment using a prototype reference implementation.
\end{itemize}

Results from our experiments show that buoyancy provides deeper insight
into workload performance in shared environments compared to traditional
heuristics in an intuitive manner that is easy to understand for
application owners and cluster administrators alike. On average, we see a $19.3\%$ better
indication of approaching bottlenecks compared to traditional heuristics.
Furthermore, we demonstrate that buoyancy may be used as a drop-in replacement for
traditional heuristics in resource management systems. 

\section{Background and Motivation}

One of the primary goals of cluster management is to make better use of already
acquired hardware. Scheduling workloads onto machines while considering 
multiple resource dimensions such as time, CPU, and memory is an instance
of the NP-hard multidimensional bin packing 
problem~\cite{christensenApproximationOnlineAlgorithms2017}.
Because of this difficulty, most modern cluster management systems, such as 
Borg~\cite{tirmaziBorgNextGeneration2020} and 
Kubernetes~\cite{carrionKubernetesSchedulingTaxonomy2022},\footnote{\url{https://kubernetes.io}} 
rely on heuristics to make decisions about workload placement and resource allocation. 

Such orchestrators operate in the domain of hardware systems, where resources are
discrete and finite. Workloads are allocated CPU cores, memory, and certain
other resources such as GPUs and other hardware-level accelerators. 
In this domain, workload resource consumption is little more than 
CPU-core utilization and memory usage numbers. 
While this is a simple approach that has worked reasonably well for many years, 
the types of workloads running in cloud environments have 
become much more diverse~\cite{qinHowDifferentAre2023}. Ranging from latency-sensitive
web services that see massive diurnal changes in resource demand, to 
memory-intensive big data or video processing workloads, to machine learning 
workloads that are both resource-intensive and in the case of inference, latency-sensitive. 
With such a heterogeneous range of workload types, it is only
natural that the resource requirements of workloads have become similarly complex.
Some workloads may be sensitive to their CPU core allocation, while others see massive
performance degradation when not enough LLC is 
available~\cite{patelCLITEEfficientQoSAware2020}. The
same may be true for other shared resources such as memory bandwidth (MBW), network bandwidth,
and disk I/O. Even though recent hardware provides support for dynamic control over shared system 
resources such as LLC and MBW, these resources are often not considered 
by the orchestrators due to the complexity they introduce~\cite{larssonHardwareLevelQoSEnforcement2025}.

\subsection{Resource Bottlenecks: The Limiting Factor}

The performance of a workload will always be determined by one or more limiting factors. 
Such factors may include resources such as compute, memory bandwidth, disk I/O, network bandwidth, and similar.
Additional factors such as incoming load, and interference from co-located workloads
may also play a role in determining performance.
Referred to as \emph{bottlenecks}, such factors vary depending on the system and
workload in question~\cite{ibidunmoyePerformanceAnomalyDetection2015}. 
In the simplest case, a single bottleneck may emerge as the limiting factor to a
workload's performance. In more complex cases, limitations in multiple resource domains can concurrently
or interdependently limit a workload's performance~\cite{malkowskiExperimentalEvaluationNtier2009}.
Furthermore, bottlenecks may change over time due to changes in load, resource contention, or other
factors. 

\begin{figure}[tbp]
    \centering
    \scalebox{.5}{
        \fontsize{14}{14}\selectfont 
        \graphicspath{{svg-inkscape/}}
        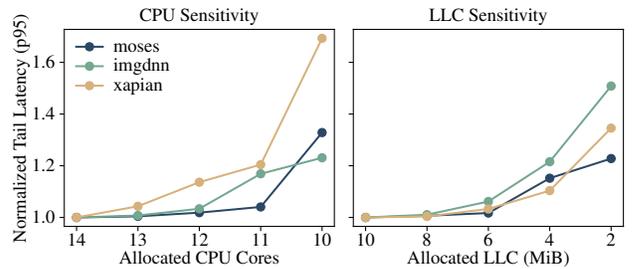
    }
    \caption{The normalized performance of different workloads as their allocation of CPU and LLC changes. 
        All other resources allocations and workload parameters are kept constant. It is clear that some 
        applications benefit more than others from the additional allocation in each domain.}
    \label{fig:bottleneck-example}
    \Description{Two graphs showing how different workloads are affected by changes in CPU and LLC allocation.}
\end{figure}

In real scenarios, this manifests as a difference in workload sensitivity to
a given resource. \figurename~\ref{fig:bottleneck-example} displays how 
the performance of workloads change as resource allocations 
are adjusted. Different workloads experience different performance degradation
when allocated less of a given resource depending on its sensitivity to the particular resource. 
It is notable that while all other resources are kept constant, the performance 
of the observed workloads decline when their LLC allocation is reduced, 
indicating that they are all sensitive to LLC to different degrees.
Even so, most modern computing orchestration platforms and resource managers 
completely disregard LLC when making scheduling and resource allocation decisions.
As yet another example of the complex interdependencies
at play, \citeauthor{chungEnforcingLastLevelCache2019}~\cite{chungEnforcingLastLevelCache2019} 
showed that in the case of LLC, increasing the LLC allocation for a workload 
can in some cases even degrade performance.

Due to such complexities, relying on CPU cores as the sole primary resource
for workload scheduling and allocation decisions is not sufficient to 
optimize for resource efficiency. 
An application owner who cares about meeting the performance targets of
their workloads may be surprised to find that increasing the only 
performance-related tuning knob available in modern infrastructure-as-a-service cloud offerings, 
CPU cores, does not improve the performance of their workload.
Alternatively, they increase the CPU allocation of their workload to the point where
sufficient amounts of other resources are ``drawn in'' by displacing other workloads.
This improves performance at the cost of utilization, stranding CPU that could otherwise have
been used by other tasks.
Providers typically employ oversubscription of hardware resources to counteract this
strategy and thus once again increase the utilization of their clusters. This can quickly
become a negative feedback loop of overprovisioning and oversubscription.
After all, a workload is only as fast as whatever bottleneck it is limited by,
and the workload may be limited by something else. 
Therefore, there is a need for another way of defining workload resource requirements.

\subsection{Observing a Trend: Intent-Driven Resource Management}

\begin{figure}[tbp]
    \centering
    \footnotesize
    \begin{tikzpicture}
	\begin{pgfonlayer}{nodelayer}
		\node [style=none] (0) at (-5.25, 0) {};
		\node [style=none] (1) at (6.25, 0) {};
		\node [style=none] (2) at (-5.25, -2.75) {};
		\node [style=none] (3) at (6.25, -2.75) {};
		\node [style=none] (4) at (1, -1) {};
		\node [style=none] (5) at (1, -2.5) {};
		\node [style=none] (6) at (6, -2.5) {};
		\node [style=none] (7) at (6, -1) {};
		\node [style=none] (9) at (-5, -1) {};
		\node [style=none] (10) at (-5, -2.5) {};
		\node [style=none] (11) at (0, -2.5) {};
		\node [style=none] (12) at (0, -1) {};
		\node [style=none] (14) at (0, -1.75) {};
		\node [style=none] (15) at (1, -1.75) {};
		\node [style=none] (16) at (-1.5, 3.5) {};
		\node [style=none] (17) at (-1.5, 0.5) {};
		\node [style=none] (18) at (4, 0.5) {};
		\node [style=none] (19) at (4, 3.5) {};
		\node [style=none] (21) at (4, 2) {};
		\node [style=none] (22) at (5, 2) {};
		\node [style=none] (23) at (5, -1) {};
		\node [style=none] (24) at (-5.25, 3.5) {};
		\node [style=none] (25) at (-5.25, 2.25) {};
		\node [style=none] (26) at (-2.75, 2.25) {};
		\node [style=none] (27) at (-2.75, 3.5) {};
		\node [style=none] (29) at (-2.75, 2.875) {};
		\node [style=none] (30) at (-1.5, 2.875) {};
		\node [style=none] (31) at (-1.5, 2.875) {};
		\node [style=none] (33) at (-5.25, 1.75) {};
		\node [style=none] (34) at (-5.25, 0.5) {};
		\node [style=none] (35) at (-2.75, 0.5) {};
		\node [style=none] (36) at (-2.75, 1.75) {};
		\node [style=none] (38) at (-2.75, 1.125) {};
		\node [style=none] (39) at (-1.5, 1.125) {};
		\node [style=none] (40) at (-4, 0.5) {};
		\node [style=none] (41) at (-4, -1) {};
		\node [style=Text] (28) at (-4, 2.875) {Intent};
		\node [style=Text] (37) at (-4, 1.125) {KPI};
		\node [style=White Text] (20) at (1.25, 2) {Intent-Driven\\Controller};
		\node [style=Text] (32) at (0.5, -0.5) {System};
		\node [style=Text] (13) at (-2.5, -1.75) {Workloads};
		\node [style=White Text] (8) at (3.5, -1.75) {Actuators};
		\node [style=none] (42) at (-5.75, 4.5) {};
		\node [style=none] (43) at (-5.75, 0.25) {};
		\node [style=none] (44) at (-2.25, 4.5) {};
		\node [style=none] (45) at (-2.25, 0.25) {};
		\node [style=Text] (46) at (-4, 4) {User-Defined};
		\node [style=none] (47) at (-1, 0.5) {};
		\node [style=none] (48) at (-1, -1) {};
	\end{pgfonlayer}
	\begin{pgfonlayer}{edgelayer}
		\draw [style=Freebox] (18.center)
			 to (17.center)
			 to (16.center)
			 to (19.center)
			 to cycle;
		\draw [style=Dashed, rounded corners=3] (3.center)
			 to (2.center)
			 to (0.center)
			 to (1.center)
			 to cycle;
		\draw [style=Freebox, fill={rgb,255: red,100; green,100; blue,100}] (4.center)
			 to (5.center)
			 to (6.center)
			 to (7.center)
			 to cycle;
		\draw [style=Freebox Green] (11.center)
			 to (12.center)
			 to (9.center)
			 to (10.center)
			 to cycle;
		\draw [style=Dashed Arrow] (15.center) to (14.center);
		\draw [style=Arrow] (21.center)
			 to [in=180, out=0] (22.center)
			 to (23.center);
		\draw [style=Freebox Yellow] (26.center)
			 to (25.center)
			 to (24.center)
			 to (27.center)
			 to cycle;
		\draw [style=Arrow] (29.center) to (31.center);
		\draw [style=Freebox Yellow] (35.center)
			 to (34.center)
			 to (33.center)
			 to (36.center)
			 to cycle;
		\draw [style=Arrow] (38.center) to (39.center);
		\draw [style=Arrow] (41.center) to (40.center);
		\draw [style=Dashed, rounded corners=3] (45.center)
			 to (43.center)
			 to (42.center)
			 to (44.center)
			 to cycle;
		\draw [style=Arrow] (48.center) to (47.center);
	\end{pgfonlayer}
\end{tikzpicture}
    \caption{A typical control-loop of an intent-driven system that manages resource allocation.}
    \label{fig:intent-driven-control-loop}
    \Description{A control loop showing how an intent-driven system manages resource allocation.}
\end{figure}
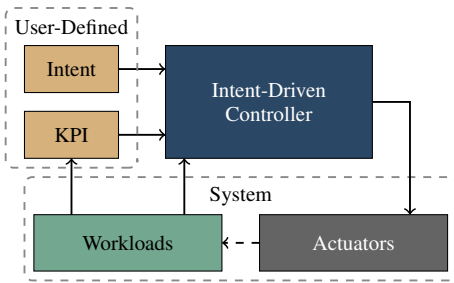

In recent years, there have been several studies proposing different takes on 
\emph{intent-driven} resource management and 
orchestration~\cite{nastic_sloc_2020, metsch_intent-driven_2023, filinis_intent-driven_2024, boutouchent_amanos_2024}.
As a concept, intent-driven management is about expressing what a system should 
achieve, rather than how it should achieve it~\cite{niemollerIntentAutonomousNetworks2022}. E.g. the requirements of a
workload may be expressed in terms of an SLO that the workload
should meet, rather than the specifics of the resources required for the workload to do so.
This opens up for more flexible and adaptive control that can adjust
to changing dynamics in the system.
The identified works leverage intent-driven approaches in systems that
set out to achieve certain goals, typically some optimized resource 
utilization~\cite{metsch_intent-driven_2023, filinis_intent-driven_2024, boutouchent_amanos_2024}.
They do this in closed-loop systems, where system inputs include workload-level 
performance metrics and SLOs. This information is then used to adjust resource
allocations to meet the demands of the workloads. Such a control loop is illustrated
in \figurename~\ref{fig:intent-driven-control-loop}.

While current intent-driven approaches are promising, their bespoke nature result in systems that
work well for only a specific purpose. Even so, the concept of expressing workload
requirements in another way than resources directly is a trend that we believe
supports a fundamental shift in how we think about workload performance and 
resource management. One that can rely on a more holistic view of resource constraints
and workload behavior, without increasing complexity for users or administrators.
However, the requirements on such a new approach are different in the eyes
of different stakeholders.

\section{Problem Statement}\label{sec:needs}

To ensure that a solution is practical and addresses real-world concerns of 
all involved stakeholders, it is important to first understand their motivations 
and requirements. In this section, we discuss the application-level, node-level, and
cluster-level needs that must be addressed by a new approach to workload and system
observability. We summarize these needs in \tablename~\ref{tab:needs}, which explicitly
maps needs to the relevant stakeholders. 
While additional stakeholders and concerns may arise in specific deployment contexts, 
we argue that the identified needs capture the primary challenges associated with 
workload and cluster observability in shared, heterogeneous environments.

\newcounter{needscounter}
\newcommand{\need}[1]{%
    \refstepcounter{needscounter}%
    \arabic{needscounter}\label{need:#1}%
}

\begin{table*}
    \centering
    \caption{Identified needs of different stakeholders.}\label{tab:needs}
    \footnotesize
    \begin{tabular}{llrl}
        \toprule
        \textbf{Stakeholder} & \textbf{Context} & \textbf{\#} & \textbf{Need} \\
        \midrule
        \multirow{2}{*}{Application Owner} & \multirow{2}{*}{Application-Level} & \need{app1} & Meet specified performance targets. \\
        & & \need{app2} & Level of sensitivity to changes in interference or load. \\ \midrule
        \multirow{6}{*}{Cluster Administrator} & \multirow{3}{*}{Node-Level} & \need{node1} & Insight into the performance of workloads. \\
        & & \need{node2} & Utilization of individual local resources. \\ 
        & & \need{node3} & Resource utilization characteristics of workloads. \\ 
        \cmidrule(l){2-4}
        & \multirow{3}{*}{Cluster-Level} & \need{cluster1} & High performance at high utilization. \\
        & & \need{cluster2} & Utilization of each node's individual resources (same as~\ref{need:node2}). \\ 
        & & \need{cluster3} & Identification of stranded resources. \\ \midrule
        \multirow{2}{*}{All} & \multirow{2}{*}{All} & \need{all1} & Intuitive and comprehensive overview of system and workload state. \\
                             & & \need{all2} & Drop-in replacement in existing systems. \\
        \bottomrule
    \end{tabular}
\end{table*}

For the \textbf{application owner} and their application-level needs, 
the primary concern is to ensure that their
workload meets its performance targets. There should also be some
notion of how sensitive the workload is to changes in interference or
load that may lead to performance degradation and thus fail to meet the performance target.
As long as these two criteria are met, the application owner does not need to care about how the
workload is scheduled or what resources it is allocated. 

The \textbf{cluster administrator's} needs are split into 
considerations on two main levels, the node-level and the cluster-level. Even so, they all
contribute to a common goal of maximizing hardware utilization and minimizing cost,
while simultaneously ensuring sufficient performance of individual workloads.

On the \textbf{node-level}, the primary concern are the workloads allocated
to the node. By understanding and leveraging the performance and resource utilization 
characteristics of every co-located workload, the node can autonomously 
adjust local resource allocations to ensure that workloads meet their performance targets,
even as workload dynamics change over time. To do this effectively, 
it is necessary to understand the workloads' utilization and saturation levels in each of
the nodes available resources. Furthermore, for feasibility in dynamic multi-tenant environments,
this characterization must be done with low overhead at runtime without 
prior workload knowledge through profiling.

In the more holistic view of \textbf{cluster-level} management, 
the cluster administrator aims to achieve high performance at high utilization while minimizing
cost across the entire cluster. Doing so requires an insight into the lifecycle
of workloads, their resource requirements, and their performance characteristics.
For example, should a workload be sensitive to a
shortage in available LLC, then it should not be scheduled on a node that is
already highly congested in that resource, even if other resources such as CPU
or memory are available. Similarly, by understanding the headroom available
in different resources on a node, it should be possible to make a prediction
whether the node can accept additional workloads without violating the SLOs of
the existing workloads. Additionally, in the event a node is overloaded, 
it should be possible to identify optimal candidates for eviction based on their
resource utilization and performance characteristics. Doing so effectively
should allow for a minimization of resource stranding, where resources become
unavailable due to high utilization of other resources locally on a node.

Finally, to support all stakeholders equally and effectively, a new approach to workload 
observability must provide an intuitive and comprehensive overview of 
system and workload states. This overview should be easily interpretable 
across roles and facilitate clear communication between application owners, 
administrators, and other involved parties. Moreover, it should integrate seamlessly 
with existing systems and workflows to ensure practical adoption in real-world environments.

The intent-driven approach to resource management has the potential of addressing all these 
needs. However, we believe the current approaches are too narrow in scope. 
Still, they indicate an important shift in how we think about workload performance 
and resource management. In fact, we believe we need a new way of thinking 
about workload performance and resource management that is not limited to a specific
system or use case, but rather a more general approach that can be applied to
a wide range of scenarios and that supports the needs in all levels of system.
From the application, to the node and the cluster.

\subsection{Solution Requirements}\label{sec:requirements}

After understanding the needs of different stakeholders, we derive a set of
requirements that a solution must fulfill to be practical and useful in real-world
scenarios. It must:

\begin{itemize}
    \item Quantify the headroom available in each resource for a workload
        given its current performance and resource usage characteristics.
    \item Adapt to changing workload dynamics over time as load
        and interference patterns change.
    \item Be generalizable to different platforms and vendors and 
        not be locked by specific hardware or features.
    \item Not require initial profiling or other prior knowledge of the workload
        in question.
    \item Be extensible to support new resources and metrics to ensure
        applicability in various domains and future systems.

\end{itemize}

\subsection{Considerations}

When considering the problem of understanding workload resource requirements, 
it is important to address a number of key concerns that arise.

\subsubsection*{Why Not Just Use the SLO with a Margin?}

Initially, this seems like a reasonable approach to address the needs of the application owner. 
The SLO is the performance target to meet, and a margin, possibly percentage based, could be added to 
provide a headroom for unforeseen increases in resource demand. However, simply
observing at the SLO does not tell the whole story. An application's performance
will always be limited by some resource, and saturating such a bottleneck typically
shows up as a \emph{knee} in the performance curve where performance starts to degrade
drastically~\cite{yaoBottleneckDetectionSolution2014}. Without an understanding of how close an application is
hitting a resource bottleneck, it is impossible to understand how much headroom
there is in the system.

\subsubsection*{Is Resource Bottleneck Detection Possible at Runtime Without Prior Profiling?}

As will become evident in later sections, detecting and predicting resource 
bottlenecks is a problem that requires a different approach depending on 
the resource in question. Accurately predicting the performance of LLC, 
for example, is a task that typically requires a deep understanding of the
workload's memory access patterns and working set size~\cite{byrne_survey_2018}, information that is in many cases 
dependent on the workload's input data and thus not available until runtime.
For example, a large language model (LLM) based inference application will have 
a working set size determined by the size of the employed model.
Gathering this information in a profiling stage is not feasible given the constraints of general-purpose cloud environments,
where it is not possible to make assumptions about workloads in advance. Therefore,
we employ a heuristic approximation of such bottlenecks based on the workload's
resource usage and performance characteristics at runtime. As we will show, 
even such approximation is sufficient to provide a useful measure of
the headroom in the system.

\section{Buoyancy Overview}

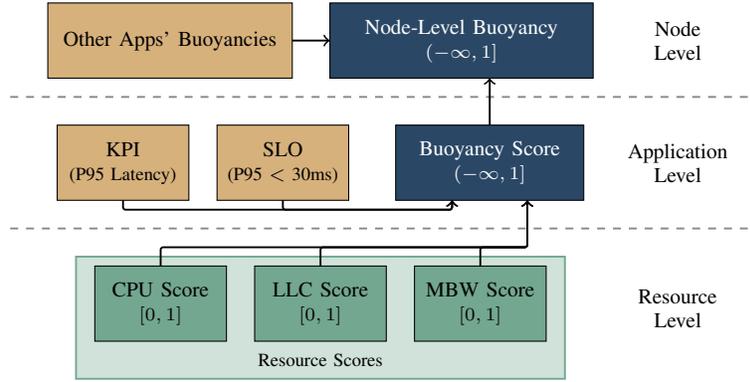
\begin{figure*}[tbp]
    \centering
    \footnotesize
    \begin{tikzpicture}
	\begin{pgfonlayer}{nodelayer}
		\node [style=none] (3) at (-7.5, 1.25) {};
		\node [style=none] (4) at (12.5, 1.25) {};
		\node [style=none] (5) at (-7.5, -2.25) {};
		\node [style=none] (6) at (12.5, -2.25) {};
		\node [style=none] (7) at (-5.25, -3.25) {};
		\node [style=none] (8) at (-5.25, -5.25) {};
		\node [style=none] (9) at (-1.75, -5.25) {};
		\node [style=none] (10) at (-1.75, -3.25) {};
		\node [style=none] (12) at (-1, -3.25) {};
		\node [style=none] (13) at (-1, -5.25) {};
		\node [style=none] (14) at (2.5, -5.25) {};
		\node [style=none] (15) at (2.5, -3.25) {};
		\node [style=none] (17) at (3.25, -3.25) {};
		\node [style=none] (18) at (3.25, -5.25) {};
		\node [style=none] (19) at (6.75, -5.25) {};
		\node [style=none] (20) at (6.75, -3.25) {};
		\node [style=none] (22) at (-6.25, 0.5) {};
		\node [style=none] (23) at (-6.25, -1.5) {};
		\node [style=none] (24) at (-2.75, -1.5) {};
		\node [style=none] (25) at (-2.75, 0.5) {};
		\node [style=none] (27) at (2.75, 0.5) {};
		\node [style=none] (28) at (2.75, -1.5) {};
		\node [style=none] (29) at (7.75, -1.5) {};
		\node [style=none] (30) at (7.75, 0.5) {};
		\node [style=none] (32) at (6.25, -1.5) {};
		\node [style=none] (33) at (6.25, -2.75) {};
		\node [style=none] (34) at (5, -2.75) {};
		\node [style=none] (35) at (0.75, -2.75) {};
		\node [style=none] (36) at (-3.5, -2.75) {};
		\node [style=none] (37) at (-3.5, -3.25) {};
		\node [style=none] (38) at (0.75, -3.25) {};
		\node [style=none] (39) at (5, -3.25) {};
		\node [style=none] (42) at (1, 3.75) {};
		\node [style=none] (43) at (1, 1.75) {};
		\node [style=none] (44) at (8, 1.75) {};
		\node [style=none] (45) at (8, 3.75) {};
		\node [style=none] (47) at (-6.5, 3.75) {};
		\node [style=none] (48) at (-6.5, 1.75) {};
		\node [style=none] (49) at (0, 1.75) {};
		\node [style=none] (50) at (0, 3.75) {};
		\node [style=none] (53) at (0, 2.75) {};
		\node [style=none] (54) at (1, 2.75) {};
		\node [style=none] (55) at (5.25, 1.75) {};
		\node [style=none] (56) at (5.25, 0.5) {};
		\node [style=none] (59) at (-2, 0.5) {};
		\node [style=none] (60) at (-2, -1.5) {};
		\node [style=none] (61) at (1.5, -1.5) {};
		\node [style=none] (62) at (1.5, 0.5) {};
		\node [style=Text] (51) at (-3.25, 2.75) {Other Apps' Buoyancies};
		\node [style=White Text] (46) at (4.5, 2.75) {Node-Level Buoyancy\\\scriptsize $(-\infty,1]$};
		\node [style=Text] (2) at (10.25, 2.75) {Node\\Level};
		\node [style=Text] (26) at (-4.5, -0.5) {KPI\\\scriptsize (P95 Latency)};
		\node [style=Text] (63) at (-0.25, -0.5) {SLO\\\scriptsize (P95 $<$ 30ms)};
		\node [style=White Text] (31) at (5.25, -0.5) {Buoyancy Score\\\scriptsize $(-\infty,1]$};
		\node [style=Text] (1) at (10.25, -0.5) {Application\\Level};
		\node [style=Text] (11) at (-3.5, -4.25) {CPU Score\\\scriptsize $[0,1]$};
		\node [style=Text] (16) at (0.75, -4.25) {LLC Score\\\scriptsize $[0,1]$};
		\node [style=Text] (21) at (5, -4.25) {MBW Score\\\scriptsize $[0,1]$};
		\node [style=Text] (0) at (10.25, -4.375) {Resource\\Level};
		\node [style=Text] (69) at (0.75, -5.75) {\scriptsize Resource Scores};
		\node [style=none] (70) at (-5.75, -3) {};
		\node [style=none] (71) at (-5.75, -6.25) {};
		\node [style=none] (72) at (7.25, -6.25) {};
		\node [style=none] (73) at (7.25, -3) {};
		\node [style=none] (74) at (-4.5, -1.5) {};
		\node [style=none] (75) at (-4.5, -1.75) {};
		\node [style=none] (76) at (-0.25, -1.5) {};
		\node [style=none] (77) at (-0.25, -1.75) {};
		\node [style=none] (78) at (4.25, -1.5) {};
		\node [style=none] (79) at (4.25, -1.75) {};
	\end{pgfonlayer}
	\begin{pgfonlayer}{edgelayer}
		\draw [style=Green EC F] (71.center)
			 to (70.center)
			 to (73.center)
			 to (72.center)
			 to cycle;
		\draw [style=Dashed] (5.center) to (6.center);
		\draw [style=Dashed] (3.center) to (4.center);
		\draw [style=Freebox Green] (10.center)
			 to (7.center)
			 to (8.center)
			 to (9.center)
			 to cycle;
		\draw [style=Freebox Green] (15.center)
			 to (12.center)
			 to (13.center)
			 to (14.center)
			 to cycle;
		\draw [style=Freebox Green] (20.center)
			 to (17.center)
			 to (18.center)
			 to (19.center)
			 to cycle;
		\draw [style=Freebox Yellow] (25.center)
			 to (22.center)
			 to (23.center)
			 to (24.center)
			 to cycle;
		\draw [style=Freebox] (30.center)
			 to (27.center)
			 to (28.center)
			 to (29.center)
			 to cycle;
		\draw [style=Arrow, rounded corners=1] (39.center)
			 to (34.center)
			 to (33.center)
			 to (32.center);
		\draw [style=Arrow, rounded corners=1] (37.center)
			 to (36.center)
			 to (33.center)
			 to (32.center);
		\draw [style=Arrow, rounded corners=1] (38.center)
			 to (35.center)
			 to (33.center)
			 to (32.center);
		\draw [style=Freebox] (45.center)
			 to (42.center)
			 to (43.center)
			 to (44.center)
			 to cycle;
		\draw [style=Freebox Yellow] (50.center)
			 to (47.center)
			 to (48.center)
			 to (49.center)
			 to cycle;
		\draw [style=Arrow] (53.center) to (54.center);
		\draw [style=Freebox Yellow] (62.center)
			 to (59.center)
			 to (60.center)
			 to (61.center)
			 to cycle;
		\draw [style=Arrow] (56.center) to (55.center);
		\draw [style=Arrow, rounded corners=1] (76.center)
			 to (77.center)
			 to (79.center)
			 to (78.center);
		\draw [style=Arrow, rounded corners=1] (74.center)
			 to (75.center)
			 to (79.center)
			 to (78.center);
	\end{pgfonlayer}
\end{tikzpicture}
    \caption{Illustration of the relationship between resources, resource scores,
        workload KPIs, and buoyancy scores.}
    \label{fig:buoyancy-parts}
    \Description{A diagram showing the relationship between resources, resource scores, workload KPIs, and buoyancy scores.}
\end{figure*}

We propose the concept of \emph{buoyancy} as a measure of resilience to changes
in load or interference in a shared resource environment. Buoyancy builds upon
a continuous analysis of workload resources and a prediction of bottlenecks to
provide a measure of headroom in the system. 
The various components of the buoyancy concept and their relationships are illustrated in
\figurename~\ref{fig:buoyancy-parts}.

In traditional physics, the measurement of buoyancy is the upward force upon an object
submerged in a fluid, based on the weight of the fluid it 
displaces~\cite{turner_buoyancy_1979}. We believe this term apt for
describing our proposed concept, as it intuitively captures the idea
of the available headroom that exists before a workload is overwhelmed by 
load or interference and therefore, sinks. \figurename~\ref{fig:buoyancy-illustration} 
displays an intuitive illustration of the concept.
In practice, buoyancy takes the form of a score that is calculated for each
workload at runtime. The buoyancy score is derived from resource scores, which
are used to quantify the resource utilization and characteristics of a workload.
Furthermore, node-level resource and buoyancy scores are defined to provide 
a holistic view of each node's resource utilization and aggregate workload performance.

To ensure applicability in a wide range of dynamic and heterogeneous general-purpose computing 
environments, buoyancy satisfies the needs outlined in \tablename~\ref{tab:needs} and fulfills the requirements detailed
in Section~\ref{sec:requirements}.

\subsection{Resource Scores}

A foundational element of the buoyancy concept is the ability to characterize 
how workloads interact with system resources. To this end, we introduce 
\emph{resource scores}, a set of runtime-computed metrics that quantify a workload’s 
utilization and sensitivity with respect to individual resources. Each workload 
is associated with a resource score per considered resource, enabling 
fine-grained insight into utilization characteristics and performance constraints.

Resource scores are values in the range $[0,1]$. A score of 0 indicates that the workload 
has no utilization of the corresponding resource and is unaffected by changes in its allocation, 
while a score of 1 implies full utilization and high sensitivity to fluctuations in availability. 
This formulation allows for intuitive interpretation: resources with higher scores are 
more likely to be performance bottlenecks under contention or interference.
Crucially however, resource scores are designed to be computed online, with minimal 
overhead and without requiring prior workload profiling or domain-specific 
knowledge. This ensures their applicability in dynamic, multi-tenant 
environments characterized by heterogeneity in both workloads and hardware. 
By comparing resource scores across resources, we can identify limiting factors 
and guide more effective resource management and scheduling decisions.

This simple definition of resource scores ensures extensibility to new resources and metrics,
and generalizability across platforms and vendors. While many resource scores will be
redefined existing metrics, some will not. The scores become a model of the workload's
interaction with the system resources, rather than a direct measurement of resource usage.
This allows for drop-in replacement if better measurement techniques or 
sensitivity models become available in the future.

\subsection{Buoyancy Score}

Building on the resource-level characterization provided by resource scores, 
we derive a \emph{buoyancy score} for each workload at runtime. The buoyancy score is a 
continuous value in the range $(-\infty, 1]$, where higher values indicate 
more available headroom before the workload becomes 
constrained by load or interference. Conversely, a score $b < 0$ signifies that 
the workload's SLO is currently violated and according to the intuition, no longer ``afloat.''

With a set of resource scores defined, we need a representation
of an application-level KPI that we can use to determine the buoyancy score
of the workload. We choose to use the performance slack as the basis for this
representation. Given a current KPI value $K_\textrm{curr} \ge 0$ and an SLO value 
$K_\textrm{SLO} > 0$, we define a KPI performance score $P$ as
\begin{equation}
    P = \frac{K_\textrm{SLO} - K_\textrm{curr}}{K_\textrm{SLO}}.
\end{equation}
In the event no KPI is defined for the workload, we consider $K_\textrm{SLO} = \infty$ 
which results in 
\[
    P = \lim_{K_\textrm{SLO}\to\infty}\frac{K_\textrm{SLO} - K_\textrm{curr}}{K_\textrm{SLO}} = 1.
\]

We now define the buoyancy score $b$ as 
\begin{equation}\label{eq:buoyancy}
    b = P \times \left(\alpha \times \left(1 - \max_{r \in R} r\right) + (1 - \alpha) \times \left( 1 - \frac{\sum_{r \in R} r}{|R|} \right) \right),
\end{equation}
where $R$ is the set of resource scores for the workload and $\alpha$ is a tunable weighting factor
between the importance of the most limiting resource and the average resource score.
This results in a buoyancy score
$-\infty < b \leq 1$, where a higher value indicates that the workload is more buoyant,
and a value $b < 0$ indicates that the workload's SLO is broken. 

The tuning factor $\alpha$ allows for adjusting the sensitivity of the buoyancy score
to the most limiting resource compared to the average of the resource scores.
A value of $\alpha = 0.7$ was experimentally selected as it provides a good balance between 
prioritizing the most limiting resource while still considering the overall
resource usage characteristics of the workload.

\begin{figure}[tbp]
    \centering
    \scalebox{.5}{
        \fontsize{14}{14}\selectfont 
        \graphicspath{{svg-inkscape/}}
        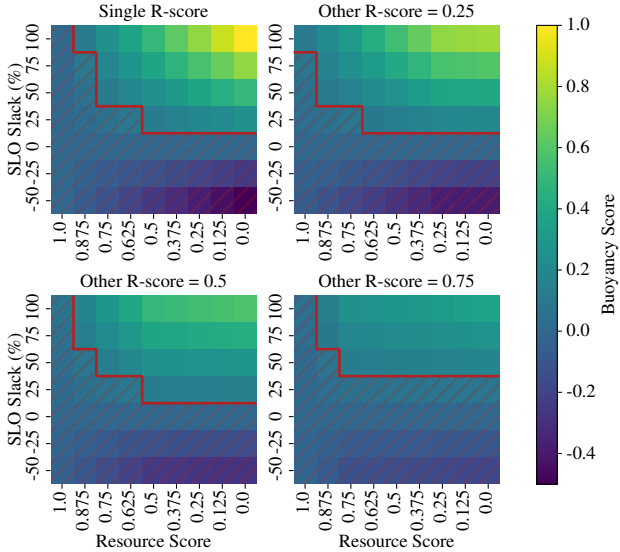
    }
    \caption{The effects of resource scores and performance slack on the buoyancy score. 
        The highlighted area shows a region where the buoyancy score $b \leq 0.1$, indicating that 
        the workload is approaching a violation. In the first case, only a single resource 
        score is considered. In the other cases, a second resource score is added and kept constant
        as indicated by the figure titles. As can be observed, even if the SLO slack is large,
        buoyancy can indicate an approaching bottleneck if resource scores are high.
    }
    \label{fig:analysis}
\end{figure}

To illustrate the effects of resource scores and performance slack on the buoyancy score, 
we present a simple visual analysis in \figurename~\ref{fig:analysis}. In this analysis,
we clearly see that buoyancy scores should allow us to earlier identify approaching
resource bottlenecks and SLO violations, compared to relying on performance slack alone.

\subsection{Node-Level Resource Scores and Buoyancy}

In addition to the workload-level resource scores and buoyancy scores,
node-level counterparts are defined to provide a holistic view of the resource
dynamics on each node. These node-level scores aggregate the characteristics
of all workloads hosted on the node, enabling system-level insights.

The node-level resource scores are defined for each resource individually, 
while adhering to the same $[0,1]$ range as the workload-level counterparts.

A node-level buoyancy score is an indication of the headroom available
to the group of workloads on the node. Therefore, the node-level buoyancy
must consider the individual workloads and their respective buoyancy scores.
The definition of a node-level buoyancy score is similar
to that of workload-level buoyancy, seen in Eq.~\ref{eq:buoyancy}. 
If $B_n$ is the set of individual buoyancy scores for all workloads hosted on node $n$, then the node-level
buoyancy $b_n$ is given by 
\begin{equation}
    b_n = \alpha \times \min_{b \in B_n} b + (1-\alpha) \times \frac{\sum_{b \in B_n} b}{|B_n|}.
\end{equation}
Here, the same tuning factor $\alpha = 0.7$ is used to balance the importance
of the most constrained workload and the average buoyancy of all workloads.

\subsection{Addressing the Needs} 

Here, we discuss how buoyancy addresses the needs on the application, node, and cluster
levels identified in Section~\ref{sec:needs}.

\subsubsection*{Workload-Level Buoyancy Needs}

For the application and its owner, the buoyancy score provides a measure of
how much headroom exists before the workload is overwhelmed by load or 
interference (needs~\ref{need:app1}, \ref{need:app2}, \ref{need:node1}, \ref{need:all1}). Together with the resource scores, 
this may serve as important information about the 
characteristics of the workload, and how it is affected by changes in the
execution environment (needs~\ref{need:node2}, \ref{need:node3}, \ref{need:cluster2}). 

\subsubsection*{Node-Level Buoyancy Needs}

The buoyancy scores and accompanying resource scores provide information
about workloads and their resource characteristics. This information can be used
by nodes to make informed resource allocation decisions at runtime (needs~\ref{need:node1}, \ref{need:node2}, \ref{need:node3}). For example,
explicitly increasing the allocation of LLC to workloads
that are sensitive to LLC and reducing the allocation to workloads that are not
may provide an overall improvement in performance and resource utilization at
little to no performance cost. 

\subsubsection*{Cluster-Level Buoyancy Needs}

Buoyancy has several use-cases even
in the context of cluster management. Firstly, buoyancy may be used as an input
to workload scheduling mechanisms, i.e. workload placement and admission control,
where the node-level buoyancy scores may 
indicate whether the node may accommodate additional workloads or not (needs~\ref{need:cluster2},~\ref{need:all2}). 
Secondly, it is possible to make workload rescheduling decisions based on 
resource scores. Should multiple workloads be sensitive to, and compete for, 
for the same limited resource, it may be beneficial to reschedule one 
of the workloads to a different node to improve overall performance (needs~\ref{need:cluster1},~\ref{need:cluster2}). 
Finally, by observing node-level resource scores, it is possible to identify underutilized (``stranded'') resources
on a node, and thus, guide scheduling and re-scheduling decisions (needs~\ref{need:cluster3},~\ref{need:all1}).

\section{Resource Scores:\\Bridging Theory and Practice}\label{sec:resource-scores}

The buoyancy concept is designed to be 
portable and extensible, allowing for the seamless integration of more advanced and 
precise algorithms should they become available without necessitating changes to the core design.

In this paper, we consider resource scores for CPU, 
last-level cache (LLC), and memory bandwidth. These represent a diverse set of 
resource types that exhibit different modelling challenges, and are commonly relevant for a wide range of workloads.
However, should other resources such as network bandwidth or disk I/O be relevant, they can be added
without altering the overall design as long as they adhere to the resource score definition.
These are considered out-of-scope, but may be modeled similarly to the included resources.

The chosen models for computing resource scores are designed to be efficient
and practical for online computation while providing sufficient accuracy and insight
to demonstrate the applicability of buoyancy. 

\subsection{CPU Score}

The CPU resource score $R_{\textrm{CPU}}$ is computed from the user space CPU time
$T_{\textrm{user}}$ and the total allocated CPU time $T_{\textrm{alloc}}$ as follows:
\begin{equation}\label{eq:cpu-score}
    R_{\textrm{CPU}} = \frac{T_{\textrm{user}}}{T_{\textrm{alloc}}}.
\end{equation}

Here, the allocated CPU time $T_{\textrm{alloc}}$ is maximum CPU time that the workload 
could have used given its current resource allocation. The motivation behind 
using only the user space CPU time and not including time in kernel space
is that in many cases, limitations in other resources such as I/O typically 
increases time spent in kernel space. This would lead to a greater CPU score, 
even though the workload is not CPU-bound. Thus, we only consider the user space 
CPU time. 

\subsection{LLC Score}

Computing a $0$ to $1$ resource score for a cache is a non-trivial task, as there is
a complex relationship between cache size, cache misses, and workload performance.
To compute an LLC resource score for a workload, we first estimate the LLC miss ratio
curve (MRC) for the workload in question. Many different methods exist 
to accomplish this~\cite{byrne_survey_2018}. However, most require profiling
or continued observation over time, both of which are unfeasible in this case.
Instead, we resort to a simple approximation of the MRC to the f.unction
\begin{equation}\label{eq:mrc}
    f(x) = ax^b,
\end{equation}
where $x > 0$ is the cache size.

To do this, we perform a linear regression in logarithmic space on the cache size
and cache misses for all available cache levels. That is, should the total
number of memory requests $N_\textrm{tot}$ be given by 
$N_\textrm{tot} = L1_\textrm{hit} + L1_\textrm{miss}$, then the cache miss ratios
$M$ for the L1, L2, and L3 caches are given by
\begin{align*}
    M_\textrm{L1} = \frac{L1_\textrm{miss}}{N_\textrm{tot}},\textrm{ }
    M_\textrm{L2} = \frac{L2_\textrm{miss}}{N_\textrm{tot}},\textrm{ and }
    M_\textrm{L3} = \frac{L3_\textrm{miss}}{N_\textrm{tot}}.
\end{align*}
The cache size $S$  for each level is given by the size of the cache in KiB.
For L1, we consider the combined size of data and instruction caches. For L3, the
current LLC allocation size is used. If no allocation is made, the total size 
of the LLC available to the workload is 
used.\footnote{The available LLC may vary depending on system hardware and its NUMA configuration, which may be less than the total package LLC size.}
We thus have the data points 
$(S_\textrm{L1}, M_\textrm{L1})$, $(S_\textrm{L2}, M_\textrm{L2})$, and
$(S_\textrm{L3}, M_\textrm{L3})$ for the linear regression.
We fit the model $f(x) = \beta_0 + \beta_1 x$ to the data points after taking the 
natural logarithm of both axis. The coefficients $\beta_0$ and $\beta_1$ are found using
the ordinary least squares method~\cite{montgomery_introduction_2021}. 
The coefficients $a$ and $b$ in our MRC approximation (Eq.~\ref{eq:mrc}) are then
given by $a = e^{\beta_0}$ and $b = \beta_1$.

With this approximation of the LLC MRC, we define the LLC resource score $R_{\textrm{LLC}}$
based on the derivative $f'(x)$ of the MRC at the current LLC allocation size $S_\textrm{LLC}$:
\begin{align}
    f'(x) &= abx^{b-1},\\ 
    R_{\textrm{LLC}} &= \min\left\{ \frac{-f'(S_\textrm{LLC}) \times \textrm{llc\_way\_size}}{M_\textrm{LLC}},\;1\right\}.
\end{align}
Assuming the cache sizes are in increasing order where
$S_\textrm{L1} < S_\textrm{L2} < S_\textrm{L3}$, the MRC is monotonically decreasing ($b < 0$). Thus,
negating the derivative gives us a positive value for all realistic cache sizes ($x > 0$).
The LLC score $R_{\textrm{LLC}}$ thus gives us an estimation of how the workload's LLC miss ratio
would change should the LLC allocation locally increase or decrease. We clamp the value to a maximum of 1
to ensure it complies with the resource score definition laid out in Section~\ref{sec:resource-scores}.
This is not an issue, as in the event the value exceeds 1, the workload is very sensitive to
changes in LLC availability and the LLC should be considered a bottleneck disregarding the initial magnitude.

This definition of the LLC score is a pragmatic approach to estimating cache sensitivity
that balances accuracy with computational efficiency. The slope of the estimated MRC
indicates how much the cache miss ratio is expected to change with changes in cache size,
thus providing a useful measure of cache sensitivity. A workload with an LLC score
close to 0 is indicative of one of two scenarios: either the majority 
of memory accesses hit and the MRC has leveled out, or the workload's working set is so large
that most memory accesses miss regardless of the available cache size. In other cases,
change to the LLC allocation would have a more pronounced effect on the cache miss ratio,
and this is reflected in a higher LLC score through an increased slope in the MRC.

\subsection{Memory Bandwidth Score}

For the memory bandwidth resource score $R_{\textrm{MBW}}$, we use the
fraction of the allocated memory bandwidth $B_\textrm{alloc}$ that the workload is currently using. 
Given the current memory bandwidth $B_\textrm{curr}$, the memory bandwidth score is defined as
\begin{equation}\label{eq:mbw-score}
    R_{\textrm{MBW}} = \frac{B_\textrm{curr}}{B_\textrm{alloc}}.
\end{equation}
In the event that no memory bandwidth allocation is made, $B_\textrm{alloc}$ is 
set to the theoretical maximum memory bandwidth of the system, computed
as the product of the memory bus width, memory speed, and number of memory channels.
While not perfect, this provides a reasonable approximation of the memory bandwidth
available to the system and workload.

\subsection{Node-Level Resource Scores}

Node-level resource scores are computed similarly to the respective workload-level
resource scores. However, instead of considering the resource usage of a single workload,
the entire system is viewed holistically. For CPU score in Eq.~\ref{eq:cpu-score}, the total user space CPU time
across all workloads is used as $T_{\textrm{user}}$, and the nodes total available CPU time 
is used as $T_{\textrm{alloc}}$.
Similarly, for the memory bandwidth score computed using Eq.~\ref{eq:mbw-score}, the total memory bandwidth
across all workloads is used as $B_\textrm{curr}$, and the theoretical maximum memory bandwidth
of the node is used as $B_\textrm{alloc}$.

Given the nature of the LLC score and its estimation of the MRC of a workload,
it does not make sense to estimate the node-level LLC score in the same manner as
workloads cache sensitivity may change significantly based on their current cache allocation 
and the experienced congestion from co-located neighbors. 
Instead, we define the node-level LLC score as the average LLC score across all workloads on the node.
This has the advantage of being simple to compute while capturing 
the overall cache sensitivity of the workloads on the node.

\section{Evaluation}

To rigorously assess the applicability of the buoyancy concept, 
we conduct a comprehensive evaluation designed to address the diverse 
stakeholder requirements outlined in Section~\ref{sec:needs}. 
We begin by detailing our experimental setup in 
Section~\ref{sec:experimental-setup}, followed by a structured set of 
evaluation questions in Section~\ref{sec:evaluation-overview}. These questions 
guide the analysis and collectively demonstrate the effectiveness and generalizability 
of buoyancy across dynamic and heterogeneous environments.

\subsection{Experimental Setup}\label{sec:experimental-setup}

In order to demonstrate the feasibility and effectiveness of the buoyancy concept, we
leverage a Kubernetes-based environment. All experiments are performed on
a single worker node while load generation, orchestration, and monitoring are
handled by separate control nodes. The hardware configuration of the worker node
is detailed in Table~\ref{tab:platform}. Telegraf\footnote{\url{https://github.com/influxdata/telegraf}}
is used to collect system level metrics, which are exported to Prometheus,\footnote{\url{https://prometheus.io}}
a time-series database used to aggregate results. 

\begin{table}[tbp]
    \centering
    \caption{Experimental platform hardware configuration.}
    \footnotesize
    \begin{tabular}{ll}
        \toprule
        \textbf{Component} & \textbf{Specification} \\ \midrule
        
        \textbf{CPU Model} & \begin{tabular}{@{}l@{}}Intel\textregistered{} Xeon\textregistered{} \\ Silver 4309Y\end{tabular} \\
        \textbf{Operating System} & Ubuntu 24.04 (kern. 6.8) \\ 
        \textbf{Sockets} & 1 \\
        \textbf{Processors (Physical / Logical)} & 8 / 16 \\
        \textbf{Frequency (Base / Max Turbo)} & 2.8 GHz / 3.6 GHz \\
        \textbf{L1 Cache per Core (Inst / Data)} & 48 KiB / 32 KiB \\
        \textbf{L2 Cache per Core} & 1280 KiB \\
        \textbf{L3 Cache (LLC)} & 12 MiB, 12 ways \\
        \textbf{Memory (Capacity / Speed)} & 128 GiB / 2666 MT/s \\
        \textbf{Memory (Channels / Bus Width)} & 4 / 8 Bytes \\
        \textbf{Storage} & 1 TB NVME SSD \\
        \textbf{Network} & 10 Gbps \\
        \bottomrule
    \end{tabular}
    \label{tab:platform}
    \Description{Experimental platform hardware configuration.}
\end{table}

\subsubsection{Workloads}

In the evaluation, we consider five different workloads that represent a range of
common cloud workload types. The workloads are summarized in Table~\ref{tab:workloads}.
Moses, img-dnn, and xapian are part of the TailBench benchmark suite~\cite{kastureTailbenchBenchmarkSuite2016} which
we adapted to run continuously in a containerized environment. Memcached and nginx are
widely used open-source applications that serve to represent common light-weight cloud native
workload types. Together, these workloads represent a diverse set of resource
requirements and performance characteristics (see Section~\ref{sec:eval-resource-scores}).

\begin{table}[tbp]
    \centering
    \caption{List of workloads used in the evaluation.}
    \label{tab:workloads}
    \footnotesize
    \begin{tabular}[c]{llll}
        \toprule
        \textbf{Ref.} & \textbf{Workload} & \textbf{Short} & \textbf{Description} \\ \midrule
        \cite{kastureTailbenchBenchmarkSuite2016} & \texttt{moses} & MO & Machine translation engine \\
        \cite{kastureTailbenchBenchmarkSuite2016} & \texttt{img-dnn} & IM & Image recognition \\
        \cite{kastureTailbenchBenchmarkSuite2016} & \texttt{xapian} & XA & Online search \\
        \cite{dormandoMemcachedDistributedMemory} & \texttt{memcached} & ME & In-memory key-value store \\
        \cite{nginxauthorsNginx2025} & \texttt{nginx} & NG & Web server \\
        \bottomrule
    \end{tabular}
    \Description{List of workloads used in the evaluation.}
\end{table}

A custom open-loop~\cite{schroeder_open_2006} load generator was developed to generate load for the workloads and
to record relevant performance metrics which are then exported to Prometheus.
The Tailbench workloads (moses, img-dnn, and xapian) are configured using
the same parameters (e.g. problem space, request distribution, etc.) as in 
the original Tailbench benchmark suite~\cite{kastureTailbenchBenchmarkSuite2016}.
The nginx workload serves $10^6$ static $1$KB HTML pages, and the load generator 
is configured to request pages according to a Zipf distribution.
Similarly, the memcached workload consists of $10^5$ keys, each previously set 
to a random value of size $200$B, which are requested
according to a Zipf distribution using \texttt{get} operations.
In our experiments, we consider the $95^{th}$ percentile latency (P95) as the primary
KPI for all workloads.

\subsection{Evaluation Overview}\label{sec:evaluation-overview}

Application buoyancy has a wide range of use cases in many different contexts. 
To ensure a broad and comprehensive evaluation that demonstrates its applicability
and its ability to fulfill the needs identified in Section~\ref{sec:needs},
we ask a series of questions that aim to cover the various aspects of application buoyancy:

\begin{itemize}
    \item \textbf{Need~\ref{need:app2}:} Are resource scores a good representaion of the sensitivity of workloads with different resource characteristics? (Section~\ref{sec:eval-resource-scores})
    \item \textbf{Needs~\ref{need:app1}, \ref{need:node1}, \ref{need:node3}, \ref{need:all1}:} Is buoyancy a good metric for identifying available headroom of workloads? (Section~\ref{sec:eval-buoyancy}) 
    \item \textbf{Needs~\ref{need:node2}, \ref{need:cluster2}, \ref{need:cluster3}:} Can buoyancy be used to identify node-level resource bottlenecks and stranded resources? (Section~\ref{sec:eval-node-bottlenecks})
    \item \textbf{Need~\ref{need:cluster1}:} Does buoyancy work as a predictor for a node's ability to take on more workloads? (Section~\ref{sec:eval-node-bottlenecks}) 
    \item \textbf{Need~\ref{need:all2}:} Is buoyancy a suitable replacement to traditional SLO-based metrics as controller input? (Section~\ref{sec:eval-slo})
\end{itemize}

Additionally, we discuss possible threats to the validity of our evaluation 
in Section~\ref{sec:threats} and discuss limitations in scope and future work in Section~\ref{sec:limitations}.

\subsection{Evaluation of Resource Scores}\label{sec:eval-resource-scores}

\begin{figure*}[tbp]
    \centering
    \begin{subfigure}[t]{0.32\textwidth}
        \centering
        \scalebox{.5}{
            \fontsize{14}{14}\selectfont 
            \graphicspath{{svg-inkscape/}}
            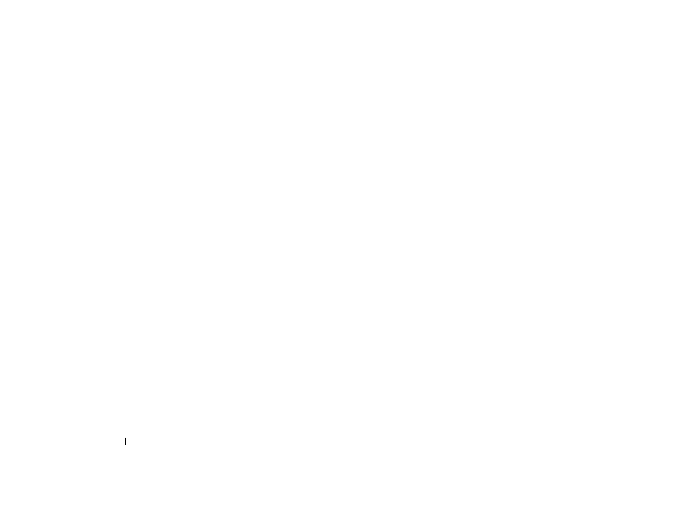
        }
    \end{subfigure}
    \hfill
    \begin{subfigure}[t]{0.32\textwidth}
        \centering
        \scalebox{.5}{
            \fontsize{14}{14}\selectfont 
            \graphicspath{{svg-inkscape/}}
            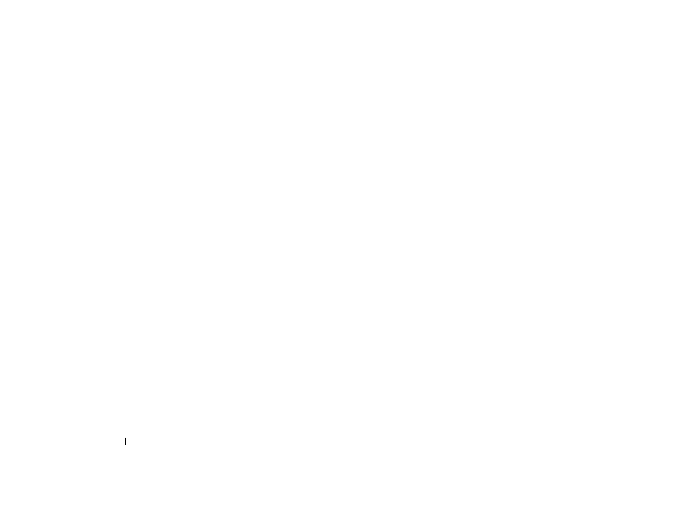
        }
    \end{subfigure}
    \hfill
    \begin{subfigure}[t]{0.32\textwidth}
        \centering
        \scalebox{.5}{
            \fontsize{14}{14}\selectfont 
            \graphicspath{{svg-inkscape/}}
            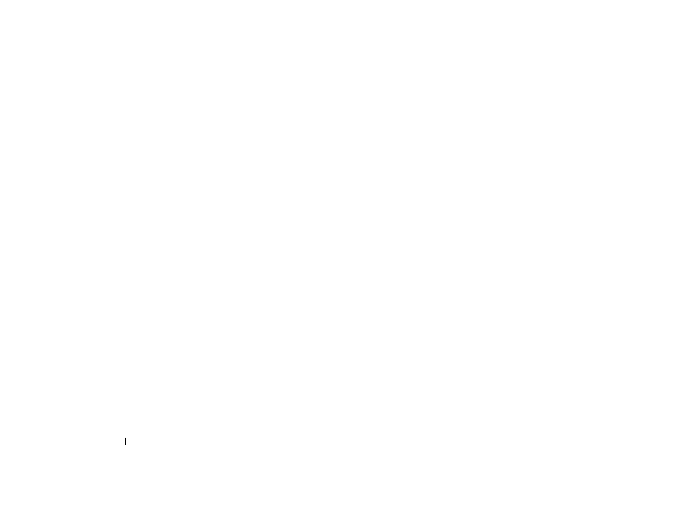
        }
    \end{subfigure}
    \caption{Resource scores for different workloads as the respective resource allocation is varied.}
    \label{fig:resourcescores}
    \Description{}
\end{figure*}

Intuitively, resource scores are a measure of a given workload's sensitivity to a particular resource. 
To verify and illustrate this intuition, we conduct an experiment where we vary the allocation of the 
considered resource as applied load is kept constant for our workloads. 
The results of individually observing the three defined resource scores, CPU, LLC and MBW, 
for our five different workloads are presented in \figurename~\ref{fig:resourcescores}.

From the results, we note that the workloads all display different resource utilization characteristics. 
In the case of the CPU score, most workloads become CPU bound at lower allocations 
for the constant applied request rate, while the nginx web server and memcached 
key-value store observe a much lower CPU score and instead show a large sensitivity to LLC at
small sizes. As pure content-serving applications, this is to be expected as they are 
lightweight in terms of CPU usage. For all applications except moses, the workloads characteristics
and the distributions of the requests make them benefit less from increased LLC allocations after a certain point.
Moses, on the other hand, sees relatively small benefits from increased LLC allocations due to 
its much larger working set size, as evident from its much larger MBW score. 

These results and the workload characteristics they imply, indicate that resource 
scores are good indicators of the resource sensitivity of workloads.

\subsection{Evaluation of Buoyancy} \label{sec:eval-buoyancy}

\begin{figure*}[tbp]
    \centering
    \scalebox{.5}{
        \fontsize{14}{14}\selectfont 
        \graphicspath{{svg-inkscape/}}
        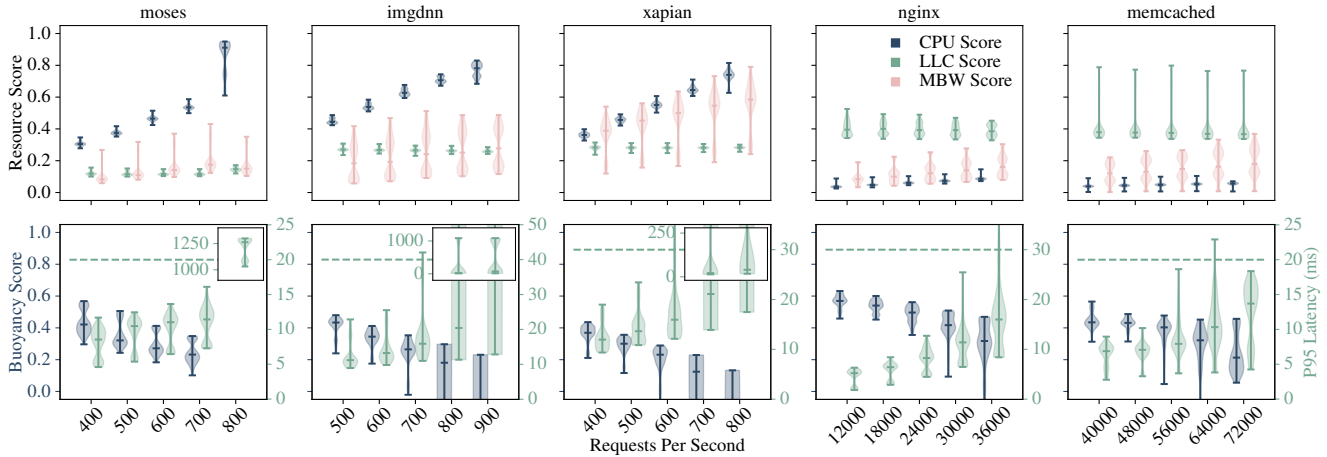
    }
    \caption{Resource scores, buoyancy score, and $95^\textrm{th}$ percentile tail 
        latency of a workload under test as the applied load is varied. 
        The workload under test is co-located with three different interfering 
        workloads, selected at random for each run of the experiment. The presented
        results are aggregated over 10 runs of the experiment. Results show that buoyancy decreases 
        before the workload experiences a sharp increase in tail latency, 
        indicating that the workload is approaching a bottleneck much earlier than is possible to 
        read from the tail latency alone. Further details are presented in \tablename~\ref{tab:headroom}.}
    \label{fig:buoyancy-test}
    \Description{A series of violin plots showing resource scores, buoyancy scores and tail latencies of different workloads as the applied load is varied.}
\end{figure*}

To answer the question of how well buoyancy represents available headroom of workloads, 
we consider a system with co-located workloads that interfere with each other. Specifically,
for each workload under test, we co-locate a combination of three different interfering workloads, selected at random out 
of the remaining four for each individual run of the experiment. All four workloads are allocated equal shares of CPU, LLC and MBW resources.
We then vary the load applied to the workload under test, while keeping the load of the interfering workloads constant.
The results of this experiment are presented in \figurename~\ref{fig:buoyancy-test}.

In the results, we observe that the resource scores of the various workloads indicate different resource sensitivities, 
as discussed in Section~\ref{sec:eval-resource-scores}. As the applied load is increased, we observe that the resource scores
increase linearly except for cases where a sharp increase in workload tail latency occurs, indicating a bottleneck. 
As the load increases in the moses, img-dnn, and xapian cases, we see that the tail latency only increases slightly before a sharp ``knee'' is observed.
Here, the buoyancy score shows a more pronounced decrease, indicating that the workload is approaching a bottleneck. 
We observe a slightly different behavior in the cases of nginx and memcached. Here, the workloads
show little changes in resource scores as the applied load is increased, indicating that
the workloads are not bottlenecked by any of the considered resources, but something else entirely. 

To further investigate this,
the median values of tail latencies and buoyancy scores from the same experiment are presented in \tablename~\ref{tab:headroom}.
In this table, we additionally observe changes in latency and buoyancy scores in percentage and log-change form. 
As we are comparing an increasing and a decreasing metric, we use log-change to compare the 
changes in a more meaningful way. There, we observe that the change in buoyancy becomes a better indicator 
of the workload's headroom than the $95^\textrm{th}$ percentile tail latency. Comparing log-changes, we have a mean 
log-change of $0.654$ for tail latency and $-0.78$ for buoyancy. If considering absolute magnitudes, 
this results in a $19.3\%$ larger actuation in buoyancy, in turn indicating a much more pronounced change.

Additionally, the changes in magnitude
in the buoyancy scores and the fact that they are approaching zero are strong suggestions that the workloads
are approaching their limits and are likely to experience performance degradation due to resource bottlenecks.

\begin{table}[tbp]
    \centering
    \caption{Median $95^\textrm{th}$ percentile tail latencies and buoyancy scores of workloads under low and high applied load. 
        These values originate from the experiment presented in \figurename~\ref{fig:buoyancy-test}. 
        The ``high'' load is the highest load applied before a knee increase in tail latency is observed.
        The log-change is calculated as
    $\ln(\textrm{high}/\textrm{low})$.}
    \label{tab:headroom}
    \footnotesize
    \begin{tabular}[c]{ll|r|r|r|r|r}
        \toprule
        \textbf{Workload} & \textbf{Ld} & \textbf{MO} & \textbf{IM} & \textbf{XA} & \textbf{NG} & \textbf{ME} \\ \midrule
        \multirow{2}{*}{RPS} & Lo & $400$ & $500$ & $400$ & $12k$ & $40k$ \\
                             & Hi & 700 & 800 & 700 & $36k$ & $72k$  \\ \midrule
        Median & Lo & 8.54 & 11.20 & 11.95 & 5.22 & 6.88 \\
        P95 (ms) & Hi & 11.43 & 20.41 & 21.10 & 15.99 & 13.71 \\ 
        \textbf{Change (\%)} & & 33.8 & 82.2 & 76.6 & 206.3 & 99.9 \\
        \textbf{Log-Change} & & 0.29 & 0.60 & 0.57 & 1.12 & 0.69 \\ \midrule
        Median & Lo & 0.42 & 0.43 & 0.37 & 0.57 & 0.43 \\
        Buoyancy & Hi & 0.23 & 0.18 & 0.12 & 0.32 & 0.21 \\
        \textbf{Change (\%)} & & $-45.2$ & $-58.1$ & $-67.6$ & $-43.9$ & $-51.1$ \\
        \textbf{Log-Change} & & $-0.60$ & $-0.87$ & $-1.13$ & $-0.58$ & $-0.72$  \\ 
        \bottomrule
    \end{tabular}
    \Description{Median $95^\textrm{th}$ percentile tail latencies and buoyancy scores of workloads under low and high applied load.}
\end{table}

With these results, we can conclude that buoyancy is a good metric for identifying the available headroom of 
workloads, as it shows a more pronounced decrease before sharp increases in tail latency 
is observed. This indicates that buoyancy can provide early warnings of approaching 
bottlenecks, before they can be observed in traditional metrics such as tail latency.

\subsection{Identification of Node-Level Resource Bottlenecks} \label{sec:eval-node-bottlenecks}

\begin{figure}[tbp]
    \centering
    \scalebox{.5}{
        \fontsize{14}{14}\selectfont 
        \graphicspath{{svg-inkscape/}}
        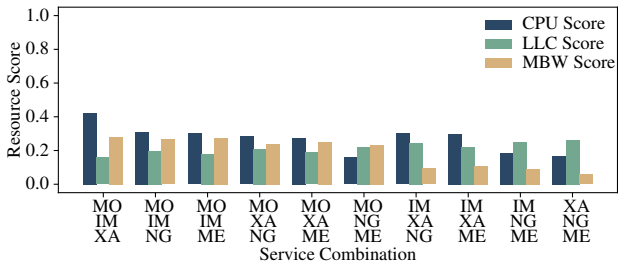
    }
    \caption{Node-level resource scores for different combinations of co-located workloads.}
    \label{fig:node_rscores}
    \Description{Node-level resource scores for different combinations of co-located workloads.}
\end{figure}

To answer whether buoyancy can be used to identify node-level resource bottlenecks and stranded resources,
we conduct an experiment where we co-locate three workloads on a single node and observe their
respective resource and buoyancy scores. The node-level resource 
scores for all $\binom 53 = 10$ combinations 
of workloads are presented in \figurename~\ref{fig:node_rscores}. 

\begin{figure}[tbp]
    \centering
    \scalebox{.5}{
        \fontsize{14}{14}\selectfont 
        \graphicspath{{svg-inkscape/}}
        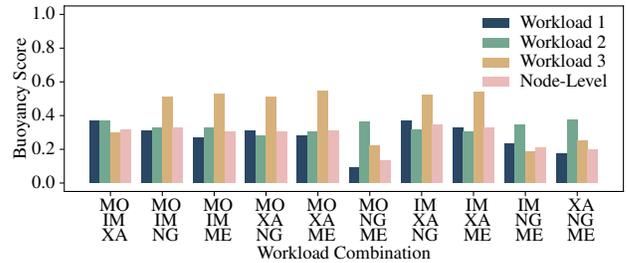
    }
    \caption{Node-level buoyancy scores for different combinations of co-located workloads.}
    \label{fig:node_buoyancy}
    \Description{Node-level buoyancy scores for different combinations of co-located workloads.}
\end{figure}

The results indicate that different combinations of workloads result in different node-level resource scores indicating 
different levels of resource contention. This is to be expected as a co-location of three CPU-intensive workloads (moses, img-dnn, and xapian) 
will contend heavily for CPU resources, whereas a better spread of resource sensitivities will generally result in less contention in all resources.
These results indicate that it is indeed possible to identify constrained and stranded resources on a node by observing the node-level resource scores. 
In addition to these results, we may also observe the node-level buoyancy score, presented in \figurename~\ref{fig:node_buoyancy}.

Here, the buoyancy scores of the individual workloads are shown in addition to the node-level buoyancy score.
When taking both the resource scores and workload SLOs into account, we see the benefit of the buoyancy score on the node-level.
Even though the resource scores indicate high contention in a resource, the node-level buoyancy score may still indicate 
a relatively large headroom
if the workloads are not close to their SLO limits. This suggests that the node can take on more workloads as long as 
the additional workloads do not significantly increase contention in the already constrained resources.
Such hints are valuable for cluster-level schedulers when making placement decisions. 

The insights of this experiment indicate buoyancy and resrouce scores can be used to 
identify node-level resource bottlenecks and stranded resources. In turn, this can be used to make an
informed assessment of a node's ability to take on more workloads, 
which is valuable for cluster-level schedulers when making placement decisions.

\subsection{Buoyancy Based Actuation} \label{sec:eval-slo}

To verify the applicability of buoyancy as a replacement for traditional SLO-based metrics as controller input,
we implemented and modified ESTHER~\cite{larssonESTHERApplicationFirstHardwareLevel2025}, a recently proposed
intent-driven resource allocator. ESTHER uses an extremum seeking approach to allocate resources
to maintain a setpoint in a workload KPI, typically tail latency. By replacing the KPI with buoyancy,
we can evaluate the ability of the controller to operate effectively using buoyancy as its input. 

We conducted an experiment similar to that which is presented in the 
ESTHER paper where the controller is maintaining 
a setpoint in the $95^\textrm{th}$ percentile tail latency of a service while the interference of
a co-located workload is varied. We used Moses as the workload under test and Phi-3, a 
large language model (LLM) as the interfering workload. Phi-3 is configured using Ollama\footnote{\url{https://ollama.com}}
in its medium variant and is prompted in a closed-loop fashion to generate text continuously using the prompt ``Why is the sky blue?''. 
During the experiments, the CPU allocation of Phi-3 is varied to create different levels of interference. As a baseline,
the experiment is first conducted with the ESTHER controller disabled.

\begin{figure}[tbp]
    \centering
    \scalebox{.5}{
        \fontsize{14}{14}\selectfont 
        \graphicspath{{svg-inkscape/}}
        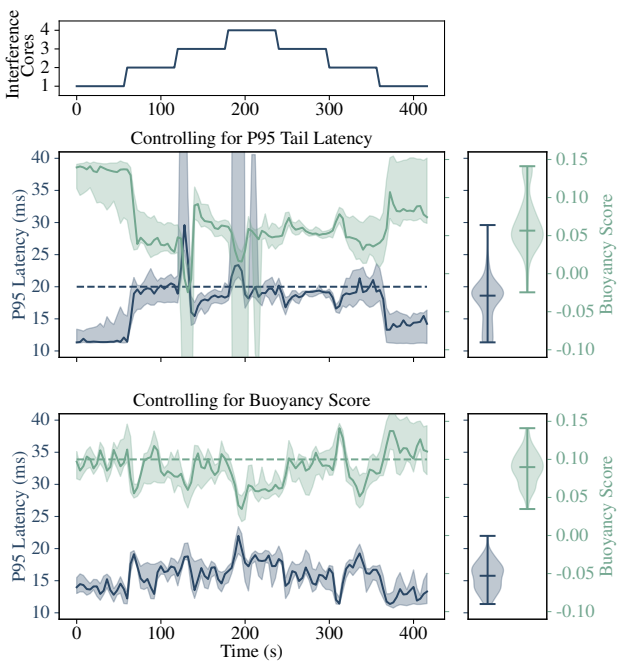
    }
    \caption{Evaluation of ESTHER~\cite{larssonESTHERApplicationFirstHardwareLevel2025}, an 
    intent-driven resource allocator, using buoyancy as a replacement for 
    traditional workload KPIs as controller input. The blue lines (left y-axis) 
    shows the p95 tail response time, while the green lines (right y-axis) represents 
    the buoyancy score. Dashed lines indicate the respective setpoints used in the 
    experiment. The number of interference cores is varied as input to the experiment 
    to induce resource contention. Each configuration is repeated 10 times; median 
    values are plotted, and shaded confidence bands represent the $10^\textrm{th}$ and $90^\textrm{th}$ percentiles. 
    The extremum seeking ESTHER controller is able to make effective use of the predictive
    nature of buoyancy to maintain the setpoint, even under varying levels of interference, 
    while the purely reactive nature of the baseline sometimes causes issues.
    Thus, buoyancy enables effective control under dynamic conditions. 
}
    \label{fig:esther}
    \Description{A graph showing the ability of ESTHER to work with buoyancy as replacement controller input instead of the workload KPI.}
\end{figure}

The results of this experiment may be observed in \figurename~\ref{fig:esther}, 
which reveal that ESTHER is able to maintain a buoyancy setpoint similarly to 
its ability to maintain a tail latency setpoint. This indicates that buoyancy 
is a feasible drop-in replacement as resource management controller input.

\subsection{Threats to Validity} \label{sec:threats}

Despite extensively covering the various aspects of resource scores and buoyancy in our evaluation,
inevitable threats to the validity of our results exist. Here, we discuss the most prominent threats,
both external and internal.

\subsubsection{External Threats to Validity}

The evaluation for this work has been performed using a cloud native technology and software stack
including technologies such as Kubernetes, Prometheus, and Telegraph. It is not
guaranteed that future versions of these technologies will behave in the same way as
in our evaluation. 

Additionally, the hardware performance counters used to derive resource scores
may not be made available in future hardware. Without hardware metrics such as
cache hit-and-miss counters, it may not be possible to derive resource scores
in the same way as presented in this work. Similarly, the implementation of
hardware performance counters may differ between hardware vendors and models,
potentially affecting the accuracy of resource scores.

\subsubsection{Internal Threats to Validity}

Care has been taken to ensure the correctness of our implementation and evaluation.
However, there are still some internal threats to the validity of our results.
Firstly, the range of workloads used in our evaluation is limited. While we have
selected workloads that cover a range of typical computing tasks, others may
exist that challenge the applicability of resource scores and buoyancy. We find this
unlikely, however, as the black-box nature of resource scores in particular
is make them applicable in all workload contexts.

Secondly, the evaluation has been performed on a single hardware platform due
to limited access to other hardware. While we have no reason to believe that
the shown applicability of resource scores and buoyancy is limited by the
specifics of the hardware platform used, it is possible other platforms behave 
in different ways. 
 
Finally, the evaluation was performed in a controlled environment where
all aspects of the system were known and could be controlled. In real-world
scenarios, the environment may be more complex and less predictable, potentially
yielding different outcomes than those presented in this work. 

\subsection{Limitations in Scope and Future Work}\label{sec:limitations}

While we have demonstrated the general applicability of application buoyancy in various contexts,
there are still many avenues for future work and improvements to be made.
Here, we discuss some limitations to the scope of this work and possible future directions.
 
In our reference implementation and evaluation, we have considered three resources, 
CPU, LLC, and MBW. However, depending on the environment and the workloads in question,
it may be necessary to consider other resources such as network bandwidth or disk I/O for a 
more complete view of the system. By defining resource scores for these resources,
they may be seamlessly integrated into the buoyancy framework. Additionally,
should better models for resource scores be developed, for example through
the use of better hardware performance counters or machine learning techniques,
they may similarly be integrated without change.

Furthermore, apart from a small scale verification, we have not demonstrated the applicability
of buoyancy in the vast range of possible application contexts. Everything from
resource management, scheduling, and autoscaling to capacity planning and
cost allocation may benefit from the insights provided by buoyancy. However, we have
demonstrated promising potential and leave it to future work to explore
these avenues.

\section{Related Work}

\subsubsection*{Intent-driven Management}

The intent-driven paradigm has taken hold in various domains of computing
in recent years. One of the most notable and explored contexts is intent-driven management
of networks~\cite{sharmaSLAManagementIntentDriven2023,abhashkumarSupportingDiverseDynamic2017,khanGenericIntentbasedNetworking2020,liIntentDrivenQoSAwareRouting2022,quTrafficEngineeringServiceOriented2020,boutouchent_amanos_2024}.
Specifically, within the software defined networks (SDN) of the backbone of the modern cellular
network, the intent-driven paradigm has become an enabler of dynamic and efficient
management at many levels~\cite{mehmoodIntentdrivenAutonomousNetwork2023,sharmaSLAManagementIntentDriven2023}
and is paving the way for the future of radio networks in 6G and beyond~\cite{wangSurveyIntentDrivenEndtoEnd2025}.

On the side of cloud orchestration and resource management,
studies have proposed intent-driven approaches to everything from resource partitioning~\cite{metsch_intent-driven_2023, larssonESTHERApplicationFirstHardwareLevel2025}
to workload scheduling~\cite{filinis_intent-driven_2024}. However, it is also arguable that more general SLO-aware resource
partitioning approaches~\cite{chenOLPartOnlineLearning2023,patelCLITEEfficientQoSAware2020} and 
scheduling systems~\cite{zhang_zeus_2021,nastic_sloc_2020} also fall under the 
umbrella of intent-driven systems as they ensure the performance targets of workloads
are met while satisfying other goals such as improved performance of co-located 
workloads~\cite{zhangLIBRAClearingCloud2021} and cluster utilization~\cite{chenAvalonQoSAwareness2019}.
These systems all leverage workload-level performance metrics
to guide resource management decisions. This insight, combined with insight into 
the complex interdependencies at play in modern computing systems, is what
allow these systems to achieve their goals.

\subsubsection*{Bottleneck Identification and Prediction}

Performance issues in computing systems are often the result of insufficient resources
available to a workload in relation to its current load and demands~\cite{ibidunmoyePerformanceAnomalyDetection2015}.
Previous works have explored the problem of identifying such performance 
anomalies~\cite{soldaniAnomalyDetectionFailure2022}, including 
root-cause analysis~\cite{xinFinegrainedRobustPerformance2024,wangCloudRangerRootCause2018,qiuCausalityMiningKnowledge2020} and prediction 
of imminent issues~\cite{xinRobustAccuratePerformance2023,denaroPredictingFailuresAutoscaling2024,pitakratHoraArchitectureawareOnline2018}.
However, such approaches are less suitable for the purpose of driving resource
management decisions as they are not lightweight enough, nor do they provide
actionable insights into the specifics of resources.

\subsubsection*{Performance in Cloud Computing Environments}

Performance monitoring systems and performance targets are an important part of everyday 
operations of shared and cloud computing environments~\cite{skarinControloverthecloudPerformanceStudy2020,silvaPerformanceEvaluationCloud2025}.
However, performance is typically considered 
in one of two separate contexts~\cite{dengCloudNativeComputingSurvey2024}. Firstly, in the
context of the applications, KPIs such as latency~\cite{zhengMultiTenantFrameworkCloud2021} and throughput~\cite{yeungHorusInterferenceAwarePredictionBased2022} 
are the main concerns. 
Secondly, in the context of the platform, other concerns such
as energy consumption~\cite{kaurKEIDSKubernetesBasedEnergy2020}, 
interference~\cite{beltreKubeSphereApproachMultiTenant2019}, and cost~\cite{chungStratusCostawareContainer2018} 
are more relevant. This separation comes naturally as the two contexts
are often handled by different stakeholders with different goals.
However, it also means that the two contexts are often treated in isolation,
which can lead to suboptimal outcomes. We believe that the buoyancy concept
bridges this gap by providing a unified view of both application-level performance
and platform-level resource utilization.

\section{Conclusions}

In conclusion, we have identified a pressing need for a more effective and 
generalizable approach to characterizing workload performance in modern, heterogeneous, 
and shared computing environments. To address this, we introduced the concept of 
buoyancy, a novel framework that captures both resource utilization and workload 
characteristics to assess performance headroom and identify bottlenecks. 
By taking into account the resource utilization 
and characteristics of workloads, buoyancy provides a measure of headroom
and identifies bottlenecks and stranded resources. These insights enable
better decision-making in resource allocation and scheduling on the application-level,
the node-level, and the cluster-level.

\printbibliography

@inproceedings{verma_large-scale_2015,
	title = {Large-scale cluster management at Google with Borg},
	doi = {10.1145/2741948.2741964},
	series = {{EuroSys} '15},
	booktitle = {Proceedings of the Tenth European Conference on Computer Systems},
	author = {Verma, Abhishek and Pedrosa, Luis and Korupolu, Madhukar and Oppenheimer, David and Tune, Eric and Wilkes, John},
	date = {2015-04-17},
}

@article{zhang_zeus_2021,
	title = {Zeus: Improving Resource Efficiency via Workload Colocation for Massive Kubernetes Clusters},
	doi = {10.1109/ACCESS.2021.3100082},
	journaltitle = {{IEEE} Access},
	author = {Zhang, Xiaolong and Li, Lanqing and Wang, Yuan and Chen, E. and Shou, Lidan},
	date = {2021-07-26},
}

@article{lorido-botran_unsupervised_2017,
	title = {An unsupervised approach to online noisy-neighbor detection in cloud data centers},
	doi = {10.1016/j.eswa.2017.07.038},
	journaltitle = {Expert Systems with Applications},
	author = {Lorido-Botran, Tania and Huerta, Sergio and Tomás, Luis and Tordsson, Johan and Sanz, Borja},
	date = {2017-12-15},
}

@article{filinis_intent-driven_2024,
	title = {Intent-driven orchestration of serverless applications in the computing continuum},
	doi = {10.1016/j.future.2023.12.032},
	journaltitle = {Future Generation Computer Systems},
	author = {Filinis, Nikos and Tzanettis, Ioannis and Spatharakis, Dimitrios and Fotopoulou, Eleni and Dimolitsas, Ioannis and Zafeiropoulos, Anastasios and Vassilakis, Constantinos and Papavassiliou, Symeon},
	date = {2024-05-01},
}

@article{metsch_intent-driven_2023,
	title = {Intent-Driven Orchestration: Enforcing Service Level Objectives for Cloud Native Deployments},
	doi = {10.1007/s42979-023-01698-0},
	journaltitle = {{SN} Computer Science},
	author = {Metsch, Thijs and Viktorsson, Magdalena and Hoban, Adrian and Vitali, Monica and Iyer, Ravi and Elmroth, Erik},
	date = {2023-03-17},
}

@article{boutouchent_amanos_2024,
	title = {{AMANOS}: An Intent-Driven Management and Orchestration System for Next-Generation Cloud-Native Networks},
	doi = {10.1109/MCOM.003.2300367},
	journaltitle = {{IEEE} Communications Magazine},
	author = {Boutouchent, Akram and Meridja, Abdellah N. and Kardjadja, Youcef and Maia, Adyson M. and Ghamri-Doudane, Yacine and Koudil, Mouloud and Glitho, Roch H. and Elbiaze, Halima},
	date = {2024-06},
}

@article{nastic_sloc_2020,
	title = {{SLOC}: Service Level Objectives for Next Generation Cloud Computing},
	url = {https://ieeexplore.ieee.org/document/9146966/},
	doi = {10.1109/MIC.2020.2987739},
	journaltitle = {{IEEE} Internet Computing},
	author = {Nastic, Stefan and Morichetta, Andrea and Pusztai, Thomas and Dustdar, Schahram and Ding, Xiaoning and Vij, Deepak and Xiong, Ying},
	date = {2020-05},
}

@book{turner_buoyancy_1979,
	title = {Buoyancy Effects in Fluids},
	isbn = {978-0-521-29726-4},
	publisher = {Cambridge University Press},
	author = {Turner, John Stewart},
	date = {1979-12-20},
}

@incollection{yaoBottleneckDetectionSolution2014,
	title = {Bottleneck Detection and Solution Recommendation for Cloud-Based Multi-Tier Application},
	booktitle = {Service-Oriented Computing},
	publisher = {Springer Berlin Heidelberg},
	author = {Yao, Jinhui and Jung, Gueyoung},
	date = {2014},
	doi = {10.1007/978-3-662-45391-9_38},
}

@article{christensenApproximationOnlineAlgorithms2017,
	title = {Approximation and online algorithms for multidimensional bin packing: A survey},
	doi = {10.1016/j.cosrev.2016.12.001},
	journaltitle = {Computer Science Review},
	author = {Christensen, Henrik I. and Khan, Arindam and Pokutta, Sebastian and Tetali, Prasad},
	date = {2017-05-01},
}

@article{carrionKubernetesSchedulingTaxonomy2022,
	title = {Kubernetes Scheduling: Taxonomy, Ongoing Issues and Challenges},
	doi = {10.1145/3539606},
	journaltitle = {{ACM} Computing Surveys},
	author = {Carrión, Carmen},
	date = {2022-12-15},
}

@inproceedings{tirmaziBorgNextGeneration2020,
	title = {Borg: the next generation},
	doi = {10.1145/3342195.3387517},
	series = {{EuroSys} '20},
	booktitle = {Proceedings of the Fifteenth European Conference on Computer Systems},
	author = {Tirmazi, Muhammad and Barker, Adam and Deng, Nan and Haque, Md E. and Qin, Zhijing Gene and Hand, Steven and Harchol-Balter, Mor and Wilkes, John},
	date = {2020-04-15},
}

@inproceedings{chungEnforcingLastLevelCache2019,
	title = {Enforcing Last-Level Cache Partitioning through Memory Virtual Channels},
	doi = {10.1109/PACT.2019.00016},
	booktitle = {28th International Conference on Parallel Architectures and Compilation Techniques ({PACT})},
	author = {Chung, Jongwook and Ro, Yuhwan and Kim, Joonsung and Ahn, Jaehyung and Kim, Jangwoo and Kim, John and Lee, Jae W. and Ahn, Jung Ho},
	date = {2019-09},
}

@misc{larssonHardwareLevelQoSEnforcement2025,
	title = {Hardware-Level {QoS} Enforcement Features: Technologies, Use Cases, and Research Challenges},
	doi = {10.48550/arXiv.2505.15542},
	author = {Larsson, Oliver and Metsch, Thijs and Klein, Cristian and Elmroth, Erik},
	date = {2025-05-21},
	eprinttype = {arxiv},
	eprint = {2505.15542 [cs]},
}

@inproceedings{qinHowDifferentAre2023,
	title = {How Different are the Cloud Workloads? Characterizing Large-Scale Private and Public Cloud Workloads},
	doi = {10.1109/DSN58367.2023.00055},
	booktitle = {53rd Annual {IEEE}/{IFIP} International Conference on Dependable Systems and Networks ({DSN})},
	author = {Qin, Xiaoting and Ma, Minghua and Zhao, Yuheng and Zhang, Jue and Du, Chao and Liu, Yudong and Parayil, Anjaly and Bansal, Chetan and Rajmohan, Saravan and Goiri, Íñigo and Cortez, Eli and Qin, Si and Lin, Qingwei and Zhang, Dongmei},
	date = {2023-06},
}

@inproceedings{patelCLITEEfficientQoSAware2020,
	title = {{CLITE}: Efficient and {QoS}-Aware Co-Location of Multiple Latency-Critical Jobs for Warehouse Scale Computers},
	doi = {10.1109/HPCA47549.2020.00025},
	booktitle = {{IEEE} International Symposium on High Performance Computer Architecture ({HPCA})},
	author = {Patel, Tirthak and Tiwari, Devesh},
	date = {2020-02},
}

@article{ibidunmoyePerformanceAnomalyDetection2015,
	title = {Performance Anomaly Detection and Bottleneck Identification},
	doi = {10.1145/2791120},
	journaltitle = {{ACM} Computing Surveys},
	author = {Ibidunmoye, Olumuyiwa and Hernández-Rodriguez, Francisco and Elmroth, Erik},
	date = {2015-07-22},
}

@inproceedings{malkowskiExperimentalEvaluationNtier2009,
	title = {Experimental evaluation of N-tier systems: Observation and analysis of multi-bottlenecks},
	doi = {10.1109/iiswc.2009.5306791},
	booktitle = {{IEEE} International Symposium on Workload Characterization ({IISWC})},
	author = {Malkowski, Simon and Hedwig, Markus and Pu, Calton},
	date = {2009-10},
}

@misc{byrne_survey_2018,
	title = {A Survey of Miss-Ratio Curve Construction Techniques},
	doi = {10.48550/arXiv.1804.01972},
	author = {Byrne, Daniel},
	date = {2018-04-05},
	eprinttype = {arxiv},
	eprint = {1804.01972 [cs]},
}

@book{montgomery_introduction_2021,
	title = {Introduction to Linear Regression Analysis},
	isbn = {978-1-119-57875-8},
	publisher = {John Wiley \& Sons},
	author = {Montgomery, Douglas C. and Peck, Elizabeth A. and Vining, G. Geoffrey},
	date = {2021},
}

@article{silvaPerformanceEvaluationCloud2025,
	title = {Performance Evaluation of Cloud Native Applications: A Systematic Mapping Study},
	doi = {10.1007/s10922-025-09937-w},
	journaltitle = {Journal of Network and Systems Management},
	author = {Silva, Francisco Airton and Trinta, Fernando A. M. and Bonfim, Michel S. and De Macedo, José Antonio F. and Rego, Paulo A. L. and Lagrota, Vinícius},
	date = {2025-10},
}

@inproceedings{larssonESTHERApplicationFirstHardwareLevel2025,
	title = {{ESTHER}: Application-First Hardware-Level {QoS}-Enforcement for Cloud Native Environments},
	doi = {10.1109/CLOUD67622.2025.00018},
	booktitle = {2025 {IEEE} 18th International Conference on Cloud Computing ({CLOUD})},
	author = {Larsson, Oliver and Metsch, Thijs and Klein, Cristian and Elmroth, Erik},
	date = {2025-07},
}

@article{dravaiPerformanceEfficiencyMultigenerational2025,
	title = {Performance and efficiency: A multi-generational benchmark of modern processors on bandwidth-bound {HPC} applications},
	doi = {10.1016/j.future.2025.107793},
	journaltitle = {Future Generation Computer Systems},
	author = {Drávai, Balázs and Reguly, István Z.},
	date = {2025-08-01},
}

@inproceedings{kastureTailbenchBenchmarkSuite2016,
	title = {Tailbench: a benchmark suite and evaluation methodology for latency-critical applications},
	doi = {10.1109/IISWC.2016.7581261},
	booktitle = {2016 {IEEE} International Symposium on Workload Characterization ({IISWC})},
	author = {Kasture, Harshad and Sanchez, Daniel},
	date = {2016-09},
}

@online{dormandoMemcachedDistributedMemory,
	title = {memcached - a distributed memory object caching system},
	url = {https://memcached.org/},
	author = {{Dormando}},
	urldate = {2025-09-16},
    date = {2025},
}

@online{nginxauthorsNginx2025,
	title = {nginx},
	url = {https://nginx.org/},
	author = {{Nginx Authors}},
	urldate = {2025-09-16},
	date = {2025},
}

@misc{niemollerIntentAutonomousNetworks2022,
	title = {Intent in Autonomous Networks},
	author = {Niemöller, Jörg and {McDonnell}, Kevin and O'Sullivan, James and Milham, Dave and Devadatta, Vinay and Machwe, Azahar and Lei, Wang and Ben Meriem, Tayeb and Mokrushin, Leonid and Yuan, Xie},
	date = {2022-08},
    howpublished = {{TM} Forum},
    note = {{IG1253}. Version 1.3.0},
}

@inproceedings{schroeder_open_2006,
	title = {Open Versus Closed: A Cautionary Tale.},
	booktitle = {Proceedings of the 3rd Conference on Networked Systems Design \& Implementation},
	author = {Schroeder, Bianca and Wierman, Adam and Harchol-Balter, Mor},
	date = {2006-05-01},
}

@article{sharmaSLAManagementIntentDriven2023,
	title = {{SLA} Management in Intent-Driven Service Management Systems: A Taxonomy and Future Directions},
	doi = {10.1145/3589339},
	journaltitle = {{ACM} Computing Surveys},
	author = {Sharma, Yogesh and Bhamare, Deval and Sastry, Nishanth and Javadi, Bahman and Buyya, Rajkumar},
	date = {2023-06-22},
}

@inproceedings{abhashkumarSupportingDiverseDynamic2017,
	title = {Supporting Diverse Dynamic Intent-based Policies using Janus},
	doi = {10.1145/3143361.3143380},
	booktitle = {Proceedings of the 13th International Conference on emerging Networking {EXperiments} and Technologies},
	author = {Abhashkumar, Anubhavnidhi and Kang, Joon-Myung and Banerjee, Sujata and Akella, Aditya and Zhang, Ying and Wu, Wenfei},
	date = {2017-11-28},
}

@inproceedings{khanGenericIntentbasedNetworking2020,
	title = {Generic Intent-based Networking Platform for E2E Network Slice Orchestration \& Lifecycle Management},
	doi = {10.23919/APNOMS50412.2020.9236962},
	booktitle = {2020 21st Asia-Pacific Network Operations and Management Symposium ({APNOMS})},
	author = {Khan, Talha Ahmed and Abbass, Khizar and Rafique, Adeel and Muhammad, Afaq and Song, Wang-Cheol},
	date = {2020-09},
}

@article{mehmoodIntentdrivenAutonomousNetwork2023,
	title = {Intent-driven autonomous network and service management in future cellular networks: A structured literature review},
	doi = {10.1016/j.comnet.2022.109477},
	journaltitle = {Computer Networks},
	author = {Mehmood, Kashif and Kralevska, Katina and Palma, David},
	date = {2023-01-01},
}

@inproceedings{liIntentDrivenQoSAwareRouting2022,
	title = {Intent-Driven {QoS}-Aware Routing Management for Flying Ad hoc Networks},
	doi = {10.1109/IWCMC55113.2022.9824766},
	booktitle = {2022 International Wireless Communications and Mobile Computing ({IWCMC})},
	author = {Li, Tong and Yang, Chungang and Yang, Lingli},
	date = {2022-05},
}

@article{quTrafficEngineeringServiceOriented2020,
	title = {Traffic Engineering for Service-Oriented 5G Networks with {SDN}-{NFV} Integration},
	doi = {10.1109/MNET.001.1900508},
	journaltitle = {{IEEE} Network},
	author = {Qu, Kaige and Zhuang, Weihua and Ye, Qiang and Shen, Xuemin and Li, Xu and Rao, Jaya},
	date = {2020-07},
}

@article{wangSurveyIntentDrivenEndtoEnd2025,
	title = {A Survey on Intent-Driven End-to-End 6G Mobile Communication System},
	doi = {10.1109/COMST.2025.3575041},
	journaltitle = {{IEEE} Communications Surveys \& Tutorials},
	author = {Wang, Yao and Yang, Chungang and Li, Tong and Ouyang, Ying and Mi, Xinru and Song, Yanbo},
	date = {2025},
}

@inproceedings{chenOLPartOnlineLearning2023,
	title = {{OLPart}: Online Learning based Resource Partitioning for Colocating Multiple Latency-Critical Jobs on Commodity Computers},
	doi = {10.1145/3552326.3567490},
	series = {{EuroSys} '23},
	booktitle = {Proceedings of the Eighteenth European Conference on Computer Systems},
	author = {Chen, Ruobing and Shi, Haosen and Li, Yusen and Liu, Xiaoguang and Wang, Gang},
	date = {2023-05-08},
}

@inproceedings{chenAvalonQoSAwareness2019,
	title = {Avalon: towards {QoS} awareness and improved utilization through multi-resource management in datacenters},
	doi = {10.1145/3330345.3330370},
	booktitle = {Proceedings of the {ACM} International Conference on Supercomputing},
	author = {Chen, Quan and Wang, Zhenning and Leng, Jingwen and Li, Chao and Zheng, Wenli and Guo, Minyi},
	date = {2019-06-26},
}

@inproceedings{zhangLIBRAClearingCloud2021,
	title = {{LIBRA}: Clearing the Cloud Through Dynamic Memory Bandwidth Management},
	doi = {10.1109/HPCA51647.2021.00073},
	booktitle = {2021 {IEEE} International Symposium on High-Performance Computer Architecture ({HPCA})},
	author = {Zhang, Ying and Chen, Jian and Jiang, Xiaowei and Liu, Qiang and Steiner, Ian M. and Herdrich, Andrew J. and Shu, Kevin and Das, Ripan and Cui, Long and Jiang, Litrin},
	date = {2021-02},
}

@article{xinRobustAccuratePerformance2023,
	title = {Robust and accurate performance anomaly detection and prediction for cloud applications: a novel ensemble learning-based framework},
	doi = {10.1186/s13677-022-00383-6},
	journaltitle = {Journal of Cloud Computing},
	author = {Xin, Ruyue and Liu, Hongyun and Chen, Peng and Zhao, Zhiming},
	date = {2023-01-14},
}

@article{denaroPredictingFailuresAutoscaling2024,
	title = {Predicting Failures of Autoscaling Distributed Applications},
	doi = {10.1145/3660794},
	journaltitle = {Reproduction Package for Article `Predicting Failures of Autoscaling Distributed Applications`},
	author = {Denaro, Giovanni and El Moussa, Noura and Heydarov, Rahim and Lomio, Francesco and Pezzè, Mauro and Qiu, Ketai},
	date = {2024-07-12},
}

@article{xinFinegrainedRobustPerformance2024,
	title = {A fine-grained robust performance diagnosis framework for run-time cloud applications},
	doi = {10.1016/j.future.2024.02.014},
	journaltitle = {Future Generation Computer Systems},
	author = {Xin, Ruyue and Chen, Peng and Grosso, Paola and Zhao, Zhiming},
	date = {2024-06-01},
}

@article{soldaniAnomalyDetectionFailure2022,
	title = {Anomaly Detection and Failure Root Cause Analysis in (Micro) Service-Based Cloud Applications: A Survey},
	doi = {10.1145/3501297},
	journaltitle = {{ACM} Computing Surveys},
	author = {Soldani, Jacopo and Brogi, Antonio},
	date = {2022-02-03},
}

@article{pitakratHoraArchitectureawareOnline2018,
	title = {Hora: Architecture-aware online failure prediction},
	doi = {10.1016/j.jss.2017.02.041},
	journaltitle = {Journal of Systems and Software},
	author = {Pitakrat, Teerat and Okanović, Dušan and van Hoorn, André and Grunske, Lars},
	date = {2018-03-01},
}

@inproceedings{wangCloudRangerRootCause2018,
	title = {{CloudRanger}: Root Cause Identification for Cloud Native Systems},
	doi = {10.1109/CCGRID.2018.00076},
	booktitle = {2018 18th {IEEE}/{ACM} International Symposium on Cluster, Cloud and Grid Computing ({CCGRID})},
	author = {Wang, Ping and Xu, Jingmin and Ma, Meng and Lin, Weilan and Pan, Disheng and Wang, Yuan and Chen, Pengfei},
	date = {2018-05},
}

@article{qiuCausalityMiningKnowledge2020,
	title = {A Causality Mining and Knowledge Graph Based Method of Root Cause Diagnosis for Performance Anomaly in Cloud Applications},
	doi = {10.3390/app10062166},
	journaltitle = {Applied Sciences},
	author = {Qiu, Juan and Du, Qingfeng and Yin, Kanglin and Zhang, Shuang-Li and Qian, Chongshu},
	date = {2020-01},
}

@inproceedings{skarinControloverthecloudPerformanceStudy2020,
	title = {Control-over-the-cloud: A performance study for cloud-native, critical control systems},
	doi = {10.1109/UCC48980.2020.00025},
	booktitle = {2020 {IEEE}/{ACM} 13th International Conference on Utility and Cloud Computing ({UCC})},
	author = {Skarin, Per and Tärneberg, William and Årzén, Karl-Erik and Kihl, Maria},
	date = {2020-12},
}

@article{dengCloudNativeComputingSurvey2024,
	title = {Cloud-Native Computing: A Survey From the Perspective of Services},
	doi = {10.1109/JPROC.2024.3353855},
	journaltitle = {Proceedings of the {IEEE}},
	author = {Deng, Shuiguang and Zhao, Hailiang and Huang, Binbin and Zhang, Cheng and Chen, Feiyi and Deng, Yinuo and Yin, Jianwei and Dustdar, Schahram and Zomaya, Albert Y.},
	date = {2024-01},
}

@article{kaurKEIDSKubernetesBasedEnergy2020,
	title = {{KEIDS}: Kubernetes-Based Energy and Interference Driven Scheduler for Industrial {IoT} in Edge-Cloud Ecosystem},
	doi = {10.1109/JIOT.2019.2939534},
	journaltitle = {{IEEE} Internet of Things Journal},
	author = {Kaur, Kuljeet and Garg, Sahil and Kaddoum, Georges and Ahmed, Syed Hassan and Atiquzzaman, Mohammed},
	date = {2020-05},
}

@inproceedings{chungStratusCostawareContainer2018,
	title = {Stratus: cost-aware container scheduling in the public cloud},
	doi = {10.1145/3267809.3267819},
	booktitle = {Proceedings of the {ACM} Symposium on Cloud Computing},
	author = {Chung, Andrew and Park, Jun Woo and Ganger, Gregory R.},
	date = {2018-10-11},
}

@inproceedings{zhengMultiTenantFrameworkCloud2021,
	title = {A Multi-Tenant Framework for Cloud Container Services},
	doi = {10.1109/ICDCS51616.2021.00042},
	booktitle = {2021 {IEEE} 41st International Conference on Distributed Computing Systems ({ICDCS})},
	author = {Zheng, Chao and Zhuang, Qinghui and Guo, Fei},
	date = {2021-07},
}

@article{yeungHorusInterferenceAwarePredictionBased2022,
	title = {Horus: Interference-Aware and Prediction-Based Scheduling in Deep Learning Systems},
	doi = {10.1109/TPDS.2021.3079202},
	journaltitle = {{IEEE} Transactions on Parallel and Distributed Systems},
	author = {Yeung, Gingfung and Borowiec, Damian and Yang, Renyu and Friday, Adrian and Harper, Richard and Garraghan, Peter},
	date = {2022-01},
}

@inproceedings{beltreKubeSphereApproachMultiTenant2019,
	title = {{KubeSphere}: An Approach to Multi-Tenant Fair Scheduling for Kubernetes Clusters},
	doi = {10.1109/CloudSummit47114.2019.00009},
	booktitle = {2019 {IEEE} Cloud Summit},
	author = {Beltre, Angel and Saha, Pankaj and Govindaraju, Madhusudhan},
	date = {2019-08},
}

\begin{IEEEbiography}[{\includegraphics[width=1in,height=1.25in,clip,keepaspectratio]{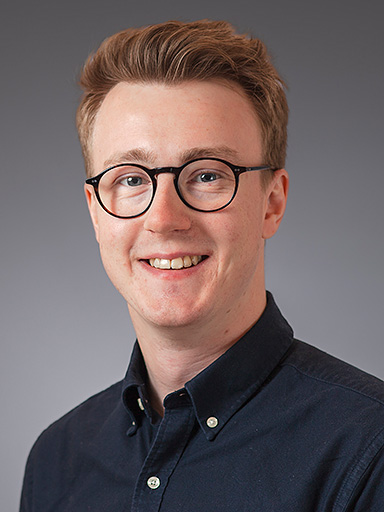}}]{Oliver Larsson}
received the M.Sc. degree in computing science and engineering from Ume\aa{} University, 
and is currently pursuing a Ph.D. at the Department of Computing Science, Ume\aa{} University, Sweden.
His main research and teaching interests lie in the intersection between cloud computing and systems performance, with a
special focus on memory subsystem optimization of previously unknown workloads in cloud native environments. 
He is the chair of the Student Council at the Royal Swedish Academy of Engineering Sciences (IVA).
Larsson is also active in the Cloud Native community, currently serving as co-organizer of the Cloud Native Ume\aa{} chapter.
\end{IEEEbiography}


\begin{IEEEbiography}[{\includegraphics[width=1in,height=1.25in,clip,keepaspectratio]{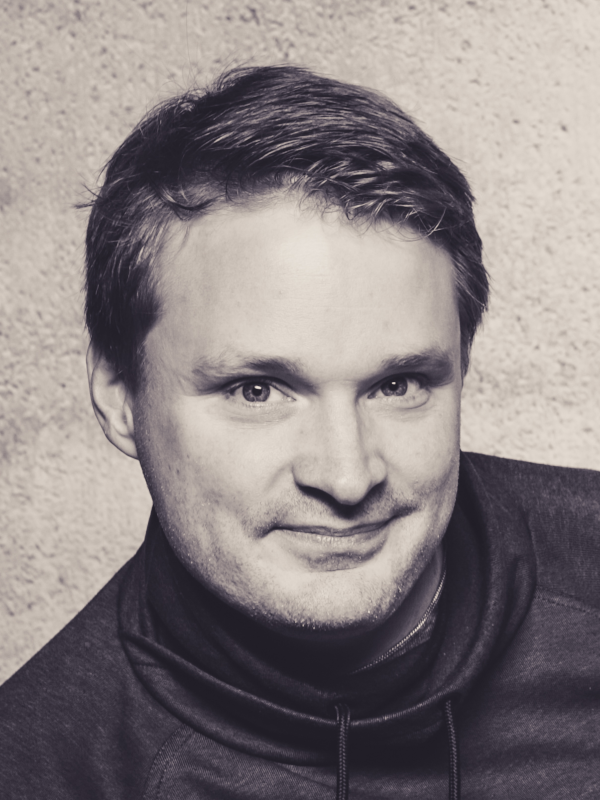}}]{Thijs Metsch}
is a distributed systems researcher and engineer with over
20 years of experience architecting, operating, and analyzing
large-scale systems across HPC, cloud, and edge environments. His work
focuses on system performance analysis, autonomic management, and the
development of intent-driven orchestration, introducing intent-based
abstractions for resource management. He co-founded and co-chaired the
Open Cloud Computing Interface (OCCI) working group defining one of
the earliest cloud computing standards and enabling interoperable,
large-scale research and industrial deployments. At Intel Labs and in
prior industry roles, he has led the development of novel
orchestration and optimization approaches applied to diverse
workloads, including cloud-native, HPC, IoT, and AI systems.
\end{IEEEbiography}


\begin{IEEEbiography}[{\includegraphics[width=1in,height=1.25in,clip,keepaspectratio]{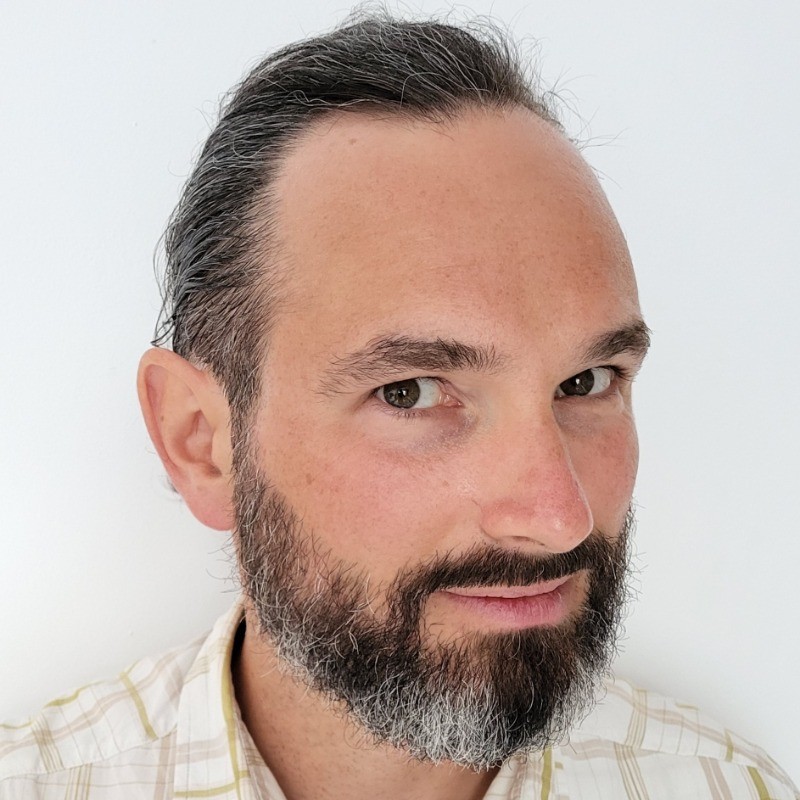}}]{Cristian Klein}
is an Adjunct Associate Professor (Docent) at Umeå University and Technical Product Owner at 
Elastisys, both located in Sweden. He works at the intersection of cloud-native architectures, 
distributed systems, software engineering and AIOps. He holds a PhD in Computing Science 
from École normale supérieure de Lyon and has over two decades of experience 
spanning academia and industry. His research focuses on adaptive and self-managing systems, 
cloud and edge computing, performance engineering, resilience in large-scale 
distributed systems and AIOps, with publications in leading venues and multiple best paper awards. 
Alongside his academic work, he has extensive industrial experience translating 
research into production-grade platforms for regulated and society-critical environments.
\end{IEEEbiography}


\begin{IEEEbiography}[{\includegraphics[width=1in,height=1.25in,clip,keepaspectratio]{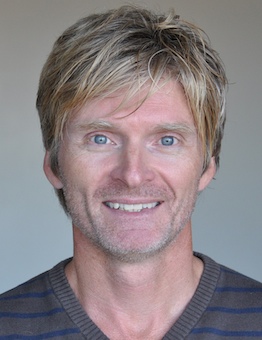}}]{Erik Elmroth}
received the Ph.D. degree in computing science from Umeå University, Umeå, 
Sweden, in 1995. He is a Full Professor of Computing Science at Umeå University, 
where he founded the Autonomous Distributed Systems Lab. He has served as 
Head and Deputy Head of the Department of Computing Science for a total of 
13 years, and as Deputy Head of a national supercomputing center for an additional 13 years.

His research spans a broad range of topics in high-performance computing, 
grid computing, and cloud infrastructure. He has received several distinctions 
for his research contributions, including the Nordea Scientific Award and the 
SIAM Linear Algebra Prize. He has also served as Chair of the Swedish National 
Infrastructure for Computing, as a member of the Swedish Research Council’s 
Committee for Research Infrastructures, and as Chair of its Expert Group on 
Scientific Infrastructures. In addition, he has authored two research strategy 
reports for the Nordic Council of Ministers, and is a Fellow of the Royal 
Swedish Academy of Engineering Sciences (IVA).
\end{IEEEbiography}

\end{document}